\documentstyle[NATO,epsfig]{crckapb}
\def\vs{\vskip 10pt}
\def\lcU{\upsilon}
\def\cU{\Upsilon}
\def\cP{{\cal P}}
\def\eg{{e.g., }}
\def\ie{{i.e., }}
\def\etal{{et al., }}

\def\etc{{etc.}}

\def\half{{\textstyle{1\over2}}}

\def\'{^{\prime}}

\def\hq{{\hat q}}

\def\C{{\cal C}}
\def\cM{{\cal M}}
\def\cW{{\cal W}}

\def\pomega{\varpi}
\def\avrg#1{{\langle #1 \rangle}}

\def\hmpc{{\, {\rm h}^{-1}~\rm Mpc}}

\def\kpc{{\rm~kpc}}
\def\mpc{{\rm~Mpc}}
\def\msun{{\,M_\odot}}

\def\spose#1{\hbox to 0pt{#1\hss}}
\def\lta{\mathrel{\spose{\lower 3pt\hbox{$\mathchar"218$}}
     \raise 2.0pt\hbox{$\mathchar"13C$}}}
\def\gta{\mathrel{\spose{\lower 3pt\hbox{$\mathchar"218$}}
     \raise 2.0pt\hbox{$\mathchar"13E$}}}
\def\ge{\mathrel{\spose{\lower 3pt\hbox{$-$}}
     \raise 2.0pt\hbox{$\mathchar"13E$}}}
\def\le{\mathrel{\spose{\lower 3pt\hbox{$-$}}
     \raise 2.0pt\hbox{$\mathchar"13C$}}}
\def\simgt{\gta}
\def\simlt{\lta}
\def\boldsymbol{\bf}
\def\be{\begin{equation}}
\def\ee{\end{equation}}
\def\bea{\begin{eqnarray}}
\def\eea{\end{eqnarray}}

\def\etal{{\it et al. }\rm}
\def\simlt{\mathrel{\hbox{\rlap{\hbox{\lower4pt\hbox{$\sim$}}}\hbox{$<$}}}}
\def\simgt{\mathrel{\hbox{\rlap{\hbox{\lower4pt\hbox{$\sim$}}}\hbox{$>$}}}}
\newcounter{parentequation}

\begin{opening}
\title{CMB ANALYSIS}

\subtitle{ }

\author{J. RICHARD BOND }
\institute{Canadian Institute for Theoretical Astrophysics\\
Toronto, ON M5S
  3H8, CANADA}
\author{ROBERT G. CRITTENDEN}
\institute{DAMTP, Centre for Mathematical Sciences, \\
University of Cambridge, 
Cambridge CB3 0WA, United Kingdom.}

\end{opening}

\runningtitle{CMB Analysis}

\begin{document}

\centerline{\bf Abstract}

\vs\noindent We describe the subject of Cosmic Microwave
Background (CMB) analysis --- its past, present and future. The
theory of Gaussian primary anisotropies, those arising from linear
physics operating in the early Universe, is in reasonably good shape 
so the focus has shifted to the statistical pipeline which
confronts the data with the theory: mapping, filtering,
comparing, cleaning, compressing, forecasting, estimating. There
have been many algorithmic advances in the analysis pipeline in
recent years, but still more are needed for the forecasts of high
precision cosmic parameter estimation to be realized. For
secondary anisotropies, those arising once nonlinearity develops,
the computational state of the art currently needs effort in all
the areas: the Sunyaev-Zeldovich effect, inhomogeneous
reionization, gravitational lensing, the Rees-Sciama effect, dusty
galaxies. We use the Sunyaev-Zeldovich example to illustrate the
issues. The direct interface with observations for these
non-Gaussian signals is much more complex than for Gaussian
primary anisotropies, and even more so for the statistically
inhomogeneous Galactic foregrounds. Because all the signals are
superimposed, the separation of components inevitably complicates
primary CMB analyses as well. 

\section{Introduction}

\begin{figure} \vspace{-0.7in}
 \centerline{ \epsfxsize=5.5in\epsfbox{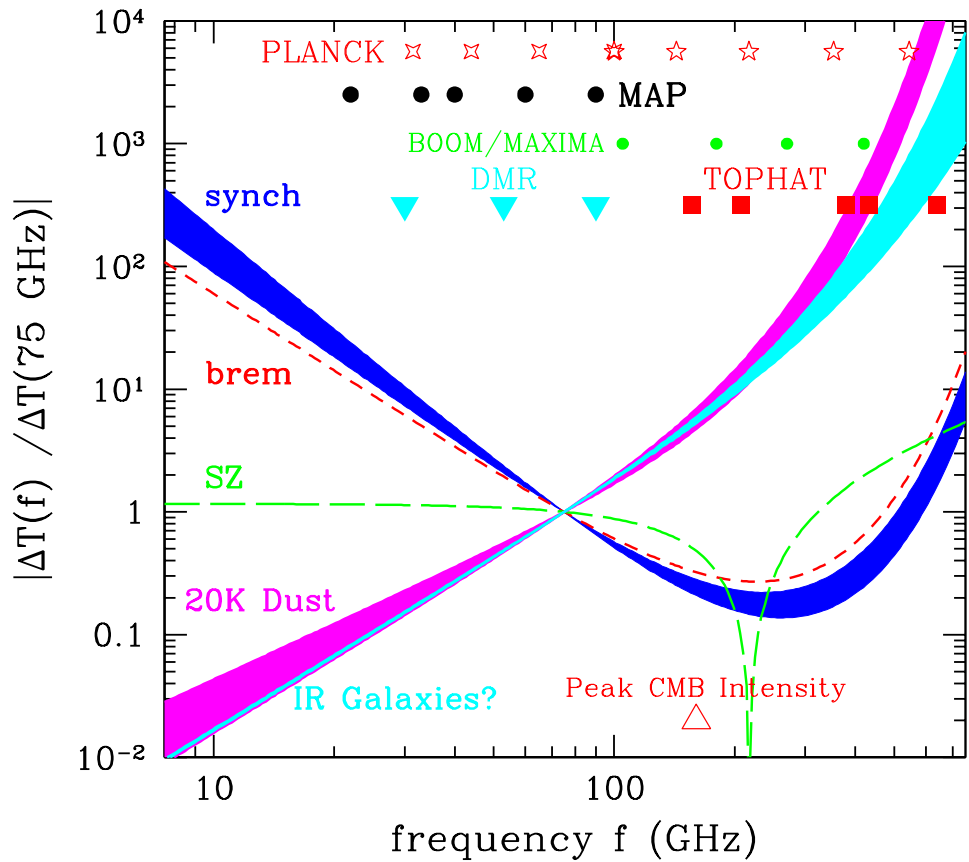}}
\caption[]{\baselineskip=10pt\small Foregrounds and secondary
anisotropies depend differently on photon frequency, the key
property for separating the components.  Plotted is the frequency
dependence of the effective thermodynamic temperature
fluctuations, normalized to their values at 75~GHz. Thus the
primary CMB fluctuations correspond to a horizontal line in this
figure, synchrotron and bremsstrahlung dominate at low
frequencies, dust at high. The bands represent a measure of our
uncertainty in the appropriate foreground spectra. The highly
distinctive shape for the Sunyaev-Zeldovich (SZ) effect is
negative at low frequencies, positive at high. The actual level of
contamination depends on the angular scale. The foreground
emission is minimal around 90~GHz, not far from where the CMB
intensity peaks. Detector frequencies for some notable experiments
are denoted by the symbols at the top. Currently, detectors below
100 GHz are HEMTs, above are bolometers. Note the wide coverage
planned for the Planck satellite which uses both. }
\label{fig:frequencyforegrounds}
\end{figure}

\begin{figure}
\vspace{-0.1in}
\centerline{ \epsfysize=6.3in\epsfbox{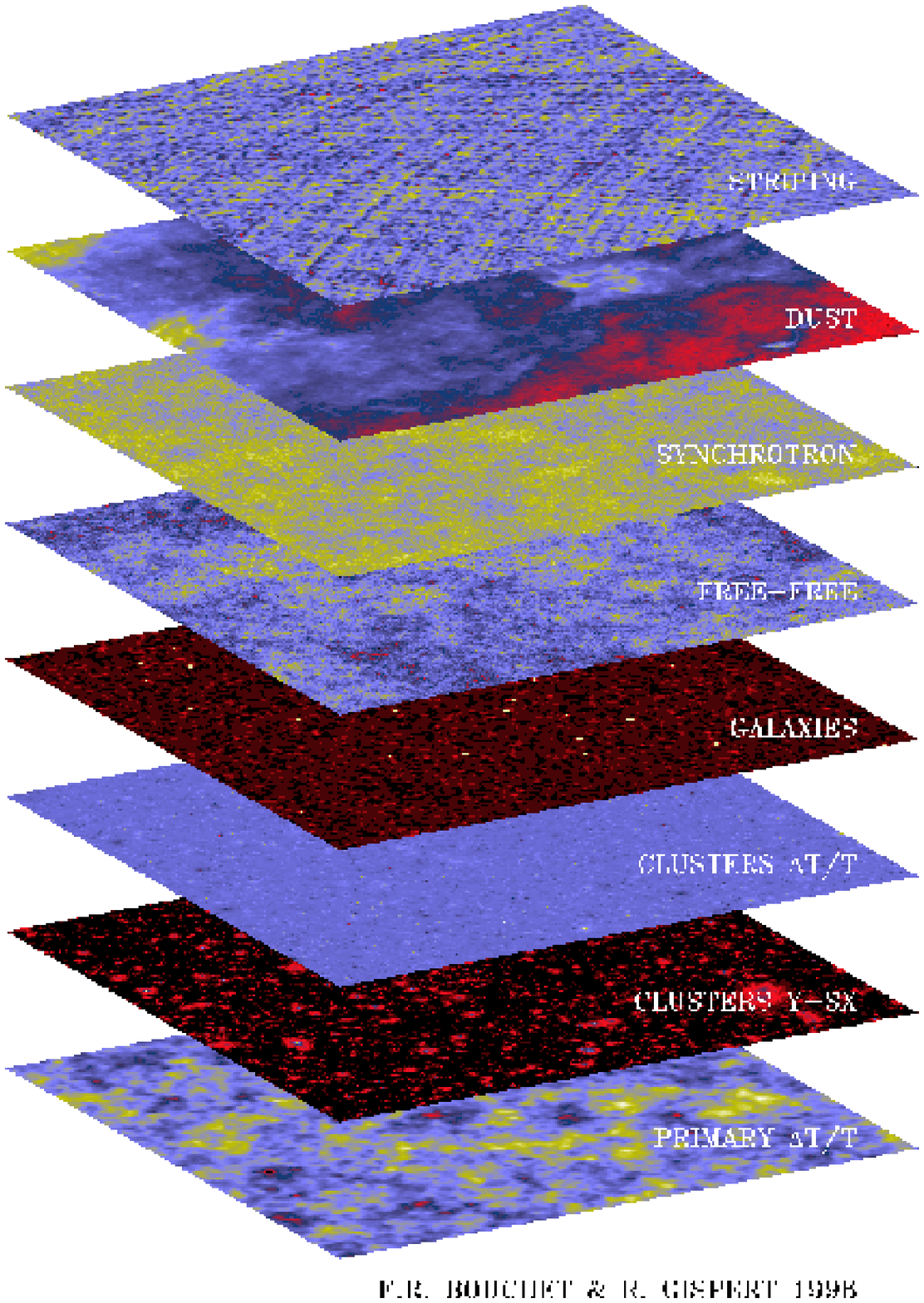} }
\caption[Sandwich]{\baselineskip=10pt\small A schematic view by
Bouchet and Gispert of many of the possible microwave foregrounds that
need to be separated from the primary CMB anisotropy pattern shown in
the $10^\circ$ map at the bottom.  These include noise effects
(striping), galactic foregrounds (dust, synchrotron and bremsstrahlung
or free-free emission) as well as secondary anisotropies
(extragalactic radio and infrared galaxies, CMB upscattering by hot
gas in clusters -- ``clusters Y-SX,'' and the Doppler
effect arising from moving clusters -- the kinetic SZ effect or 
``clusters $\Delta T/T$'').
Each of these has a unique temperature pattern on the sky, and all but
the kinetic SZ effect have a different spectral signature
(Fig.~\ref{fig:frequencyforegrounds}).}
\label{fig:spatialforegrounds}
\end{figure}

\noindent {\bf 1.1 What is CMB Analysis?}  The subject we call ``CMB
analysis" is a blend of basic theory, simulation and statistical
data analysis. The goal of CMB analyzers is to extract 
the physics of the various signals that contribute to the data. CMB
analyzers may therefore be theorists or experimentalists.
Depending upon how advanced the state-of-the-art is, the relevant
analysis topic may lean more towards the
theory/simulation side or more towards the statistical analysis
and Monte Carlo simulation side. In the CMB field, we adopt a
theorist's distinction between {\it primary}, {\it secondary} and
{\it foreground} anisotropies, though
of course the observations
do not know of this distinction. The primary ones are those we can
calculate using linear perturbation theory (or, in the case of
cosmic defects, with linear response theory).  This covers the crucially
important epoch of photon decoupling near redshift 1100, and even
until the current time on very large scales, and to redshifts of a
few on intermediate scales. Secondary anisotropies are those
associated with nonlinear phenomena, either calculable via
weakly nonlinear perturbation theory, semi-analytic methods or
by more direct simulation of nonlinear patterns.  Gravitational lensing
effects on the CMB, quadratic nonlinearities and the kinetic
Sunyaev-Zeldovich effect associated with Thomson scattering from
flowing matter, reionization inhomogeneities, the thermal
Sunyaev-Zeldovich effect associated with Compton upscattering from
hot gas, the Rees-Sciama effect associated with nonlinear
potential wells, all come under this heading. We also
traditionally call emission by dust in high redshift galaxies a
secondary process, and even emission from extragalactic radio
sources. On top of this, there are various foreground emissions from dust and
gas in our Milky Way galaxy --- signals which are
nuisances to the primary CMBologist but of passionate interest to
interstellar medium astronomers. Fortunately, most of the
secondary and foreground signals have very different dependencies
on frequency (Fig.~\ref{fig:frequencyforegrounds}), and rather
statistically distinct sky patterns
(Fig.~\ref{fig:spatialforegrounds}). Collateral information from
non-CMB observations can also be brought to bear to unravel the
various components.

\vs \noindent {\bf 1.2 Primary Theory and Current Data}: We think we
know how to calculate the primary signal in exquisite detail. The
fluctuations are so small at the epoch of photon decoupling that
linear perturbation theory is a superb approximation to the exact
non-linear evolution equations. Intense theoretical work over
three decades has put accurate calculations of this linear
cosmological radiative transfer problem on a firm footing, and
there are speedy, publicly available and widely used codes for
evaluation of anisotropies in a wide variety of cosmological
scenarios, e.g. ``CMBfast'' \cite{cmbfast}.
These have been further modified by many
different groups to attack even more structure formation models.

\begin{figure}
\vspace{-0.3in}
\centerline{\epsfxsize=5.5in\epsfbox{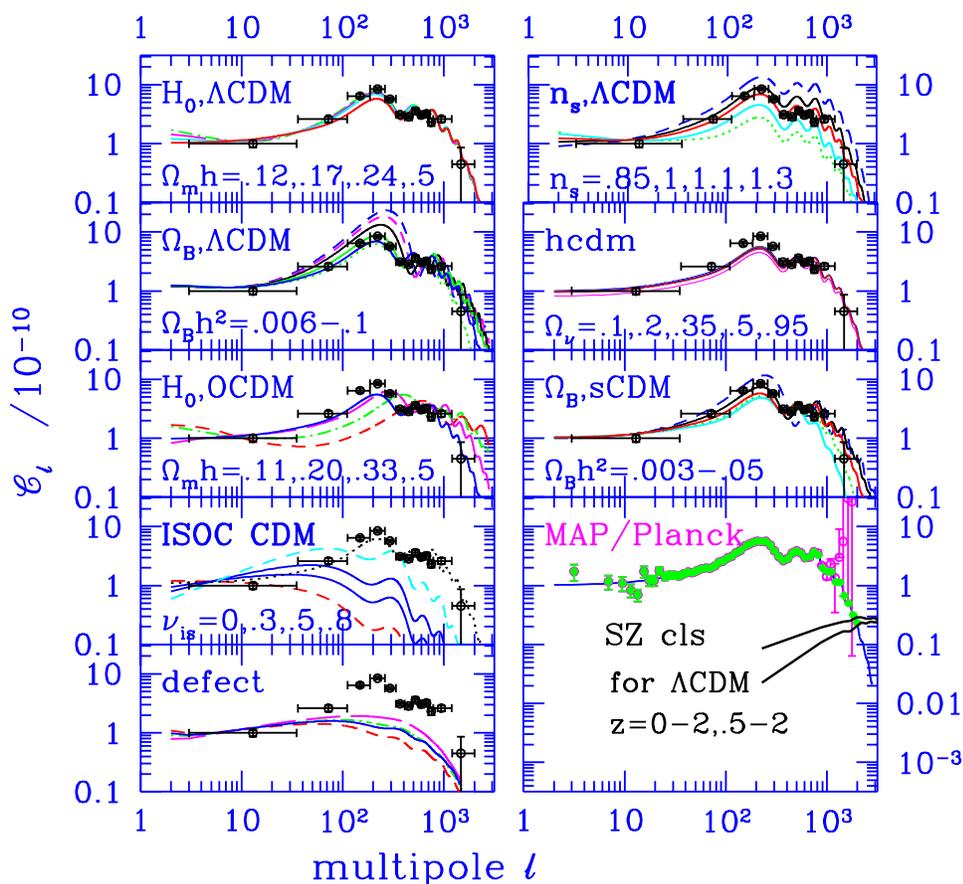} }
\vspace{-0.5in}
\caption{\baselineskip=10pt\small The ${\cal C}_\ell$ anisotropy
bandpower data, including the recent balloon-borne and interferometry
data, compressed to 13 bands using the methods of
{\protect\cite{bjk98}} are compared with various ${\cal C}_\ell$ model
sequences, each for universes with age 13 Gyr (left to right): (1)
untilted flat $\Lambda$CDM sequence with $H_0$ varying from 50 to 90,
$\Omega_{\Lambda}$ from 0 to 0.87; (2) $n_s$ varying from 0.85 to 1.3
for an $H_0=70$ ($\Omega_{\Lambda}$=.66) $\Lambda$CDM model -- dotted
is 0.85 with gravity waves, next without, upper dashed is 1.25,
showing visually why $n_s$ is found to be nearly unity; (3)
$\Omega_B{\rm h}^2$ varying from 0.0063 to 0.10 for the $H_0$=70
$\Omega_{\Lambda}$=.66 $\Lambda$CDM model; (4) neutrino fractions
$\Omega_{hdm}/\Omega_m$ varying from 0.1 to 0.95 for an $H_0$=50
$\Omega_{\Lambda}$=0 sequence with $\Omega_{m}$=1; (5) $H_0$ from 50
to 65, $\Omega_{k}$ from 0 to 0.84 for the untilted oCDM sequence,
showing the strong $\ell$-shift of the acoustic peaks with $\Omega_k$;
(6) $\Omega_B{\rm h}^2$ varying from 0.003 to 0.05 for the $H_0=50$
$\Omega_{\Lambda}$=0 sCDM model; (7) an isocurvature CDM sequence with
positive isocurvature tilts ranging from 0 to 0.8; (9) sample defect
${\cal C}_\ell$'s for textures, \etc \ from {\protect\cite{pst97}} --
cosmic string ${\cal C}_\ell$'s from {\protect\cite{allen97}} are
similar and also do not fare well compared with the current data. The
bottom right panel is extended to low values to show the magnitude of
secondary fluctuations from the thermal SZ effect for the $\Lambda$CDM
model. The kinematic SZ ${\cal C}_\ell$ is significantly lower.  Dusty
emission from early galaxies may lead to high signals, but the power
is concentrated at higher $\ell$, with a weak tail because galaxies
are correlated extending into the $\ell \lta 2000$ regime. Forecasts
of how accurate ${\cal C}_\ell$ will be determined for an sCDM model
from MAP (error bars growing above $\ell \sim 700$) and Planck (small
errors down the ${\cal C}_\ell$ damping tail) are also shown.}
\label{fig:newtnatocl9}
\end{figure}

The simplest and thus least baroque versions of inflation theory predict
that the fluctuations from the quantum noise that give rise to 
structure form a Gaussian random field. Linearity
implies that this translates into temperature anisotropy patterns
that are drawn from a Gaussian random process and which can be
characterized solely by their power spectrum. The emphasis is
therefore on confronting the theory with the data in power
spectrum space, as in Fig.~\ref{fig:newtnatocl9}. The panels show
how the power spectrum ${\cal C}_\ell$
responds as various cosmological parameters
are individually changed.

Until a few years ago, the emphasis was on fairly restricted parameter
spaces, but now we consider spaces of much larger
dimension. These are used for forecasts of how well proposed
experiments can do, and for the first round of data from a 
sequence of high-precision experiments covering large areas of sky:
ground-based single dishes and interferometers; balloons of short and 
long duration (LDBs, $\sim$ 10 days) and eventually of ultra-long
duration (ULDBs, $\sim$ 100 days); NASA's MAP satellite in 2001 and
ESA's Planck Surveyor in 2007. The lower right panel of
Fig.~\ref{fig:newtnatocl9} gives a forecast of how well we think that
the two satellite experiments can do in determining $C_\ell$ if
everything goes right. This rosy prognosis should be compared with the
compressed bandpowers shown in the other panels estimated from all of
the current data: DMR and some 19 other pre-conference ``prior''
experiments, TOCO \cite{miller00}, unveiled about the time of the
conference, and four that appeared after the conference, the
Boomerang flights (the LDB flight 
\cite{debernardis00,boom01} and its North America test
flight \cite{mauskopf00})  
the Maxima-I~\cite{hanany00,maxima01} flight, the first results from CBI,
the Cosmic Background Imager~\cite{cbi00}, and from DASI, the Degree
Angular Scale Interferometer~\cite{dasi01}. Although new methods have
been applied to these data sets, much was in place and some algorithms
were well tested on previous data, in particular DMR, SK95, MSAM
and QMAP. 

Although we sketch the current state of the art in CMB analysis in
this paper, note that this is a very fast moving subject.
A large number of researchers in a handful of groups are working 
on a broad range of issues, often as members of the experimental
teams associated with the increasingly complex experiments. A
snapshot of the state two years ago and references to the relevant
papers was given in \cite{bcjk99}, but many advancements have been made
since then. For example, the MAP plan for analysis is described in
\cite{hinshaw00}. One issue that still needs to be addressed 
is that the strong simplifying assumptions of signal
and noise Gaussianity cannot be correct in
detail, and could yield misleading results.

\vs \noindent {\bf 1.3  Primary Parameter Sets for Gaussian
Theories}: A ``minimal" inflation-motivated parameter set involves
about a dozen parameters, \eg $\{\omega_b,\omega_{cdm}$,
$\omega_{hdm},\omega_{er}$, $\Omega_{k},\Omega_\Lambda$, $\tau_C$,
$\sigma_8 \, , n_s$, $\tilde{r}_t,n_t\}$.  For many species we use 
$\omega_j \equiv \Omega_j {\rm h}^2$ rather than $\Omega_j$ since
it directly gives the physical density of the particles rather
than as a ratio to the critical density: thus, 
$\omega_b$ for baryons, $\omega_{cdm}$ for cold dark matter, $\omega_{hdm}$
for hot dark matter (massive but light neutrinos), and
$\omega_{er}$ for relativistic particles present at decoupling
(photons, very light neutrinos, and possibly weakly interacting
products of late time particle decays). The total non-relativistic
matter density is $\omega_m \equiv
\omega_{hdm}+\omega_{cdm}+\omega_b$.

The curvature energy relative to the closure density  is $\Omega_k
\equiv 1-\Omega_{tot}$ and $\Omega_\Lambda$ is the energy relative
to closure in the cosmological constant. $\Omega_\Lambda$ can also
be interpreted as a vacuum energy, a sector which may be more
complex and require more parameters. For example, a scalar field which
dominates at late times, ``quintessence", has potentials and other
interactions which must be set. A simple phenomenology of quintessence has
added to $\Omega_Q$ an average equation of state
pressure-to-density ratio $w_Q\equiv \bar{p}_Q/\bar{\rho}_{Q}$.

Another factor crucial for predicting the CMB anisotropies is
$\tau_C$, the Compton depth to Thompson scattering
from now back through the period when the universe was reionized
by early starlight.

To characterize the initial fluctuations, amplitudes and tilts are
required for
the various fluctuation modes present. For adiabatic
scalar perturbations, the tilt is $n_s$ and the 
amplitude is most appropriately the
overall power in curvature fluctuations, but is often replaced
by $\sigma_8^2$, a density bandpower on the scale of clusters
of galaxies, $8 \hmpc$.  For tensor modes
induced by gravity waves, the amplitude parameter is
some measure of the ratio of gravity wave to scalar curvature
power, $\tilde{r}_t$; usually a ratio of ${\cal C}_\ell$'s is
used, \eg at $\ell$=2 or 10. Inflation theories often give
relationships between the tensor tilt $n_t$ and $\tilde{r}_t$
which can be used to  reduce the parameter space \eg \cite{bh95}.
Parameters for the amplitude and tilt of  scalar isocurvature
modes could also be added to the mix. The characterization of the modes
present in the early universe by amplitude and tilt could be
expanded to include parameters describing the change of the tilts
with scale ($dn_s /d\ln k$), the change of that
change, and so on. Relatively full functional freedom in $n_s(k)$
is possible in inflation models, and substantial freedom also
exists for $n_t(k)$.

\vs \noindent {\bf 1.4 Well and Poorly Determined Parameter
Combinations}: In Fig.~\ref{fig:newtnatocl9}, we compare the
compressed bandpowers with ${\cal C}_\ell$ sequences as
$\omega_b$, $\Omega_{tot}$, $\Omega_\Lambda$, $\omega_{hdm}$ and
$n_s$ are individually varied. These show the discriminatory power
of the current data. Fixing the age at 13 Gyr as we do here 
defines a specific relation between $\omega_k$,
$\omega_\Lambda$ and $\omega_m$. From these figures, we expect
that four can be determined reasonably well ($\omega_b,
\Omega_{tot},n_s$ and the overall amplitude). This is borne out in
the full study with the real data, modulo parameter correlations
which mean the best determined quantities are linear combinations
of parameters, or parameter eigenmodes~\cite{eb98,bet,bh95}. For
example, it is actually a combination of $\Omega_{tot}$ and
$\Omega_\Lambda$ which is well determined, but it happens to be
$\Omega_{tot}$--dominated. Another orthogonal combination of the
two will remain very poorly determined no matter how good the CMB
data gets \cite{eb98}, an example of a near-degeneracy among
cosmological parameters.  Other near-degeneracy examples are less
severe: \eg $\Omega_{hdm}/(\Omega_{cdm}+\Omega_{hdm})$ is not well
determined by current data, but could be by Planck.

Accompanying the idealized bandpower forecasts are predictions for
how well the cosmological parameters in inflation models can be
determined after integrating, \ie marginalizing, the likelihood
functions over all other parameters. In one exercise that allowed
a mix of nine cosmological parameters to characterize the space of
inflation-based theories (the 11 above with $\omega_{er}$ fixed
and $n_t$ slaved to $\tilde{r}_t$), COBE was shown to determine
one combination of them to better than 10\% accuracy, LDBs and MAP
could determine six, and Planck seven \cite{bet}. All
currently planned LDBs could also get two combinations to 1\%
accuracy, MAP could get three, and Planck five. With the current
data, one can forecast four to 10\% accuracy, which is
actually borne out in detailed computations with the data.

Two panels in Fig.~\ref{fig:newtnatocl9} show what happens when
two classes of pure isocurvature models are considered, one a Gaussian
isocurvature CDM model with the spectral index allowed to tilt
arbitrarily, another a set of non-Gaussian cosmic defect models.
Neither fare well with the current data. Primary
anisotropies in defect theories are more complicated to calculate
because non-Gaussian patterns are created in the phase transitions
which evolve in complex ways and which require large scale
simulations. The non-Gaussianity means that the comparison with
the data should be done more carefully, without the Gaussian
assumption that goes into the bandpower estimations.

\vs \noindent {\bf 1.5 Current CMB and CMB+LSS Constraints}:
Although the purpose of this paper is to describe CMB analysis
techniques, the comparison of the radically-compressed current
bandpowers with the models in Fig.~\ref{fig:newtnatocl9} invite a
brief description of the results. Explicit numbers are those
derived in \cite{boom01} for the ``minimal'' inflation-motivated
7-parameter set for the combination of DMR and Boomerang, with a
weak prior probability assumption on the Hubble parameter 
($0.45 <h < 0.90$) 
and age
of the Universe ($>10$ Gyr.) Including
all other current CMB data as well gives very similar estimates
and errors, reflecting the consistency between the older,
less statistically significant data and
the new experiments. (DMR+DASI numbers~\cite{dasi01} are quite
close to DMR+Boomerang numbers, since the spectra are similar.  For a more
complete description, see
\cite{lange00,jaffe00,bmaxiboom00,boom01,dasi01,maxima01}.)

\noindent {\bf 1.5.1 CMB-only Estimates}: As the upper right panel
of Fig.~\ref{fig:newtnatocl9} suggests, the primordial spectral
tilt $n_s$ is well determined using CMB data alone: $n_s$ =
$0.97^{+.10}_{-.08}$. This is rather encouraging for the nearly scale
invariant models preferred by inflation theory. The figure also
shows that constraints on $\Omega_\Lambda$ will not be very good,
but the strong dependence of the position of the acoustic peaks on
$\Omega_k$ means that it is better restricted:
$\Omega_{tot}\approx 1.02 \pm 0.06 $. The baryon abundance is also
well determined, $\Omega_b h^2= 0.022\pm 0.004$, near the
$0.019\pm 0.002$ estimate from deuterium observations in quasar
absorption line spectra combined with Big Bang Nucleosynthesis
theory.

The isocurvature CDM models with tilt $n_{is}>0$ and the
isocurvature defect models ({\it e.g.}  strings and textures)
\cite{pst97,allen97} shown in Fig.~\ref{fig:newtnatocl9} clearly
have more difficulty with the CMB bandpowers.

There is a region of the spectrum that nominally seems to be out
of reach of CMB analysis, but is actually partly accessible
because ultralong waves contribute gentle gradients to our CMB
observables. For example, the size of compact spatial manifolds
has been constrained; \cite{bps00} find for flat equal-sided
3-tori, the inscribed radius must exceed $1.1 \chi_{lss}$ from DMR
at the $95\%$ confidence limit, where $\chi_{lss}$ is the distance
to the last scattering surface. For asymmetric 1-tori, the
constraint is weaker, $> 0.7 \chi_{lss}$, but still reasonably
restrictive. It is also not as strong if the platonic-solid-like
manifolds of compact hyperbolic topologies are considered, though
the overall size of the manifold should as well be of order the
last scattering surface radius to avoid conflict with the large
scale power seen in the COBE maps \cite{bps00}.

\noindent {\bf 1.5.2 CMB+LSS Estimates}: We have always combined
CMB and LSS data in our quest for viable models. For example, DMR
normalization determines $\sigma_8$ to within 7\%, and comparing
with the $\sigma_8 \sim 0.55\Omega_{m}^{-0.56} $ target value
derived from cluster abundance observations severely constrains
the cosmological parameters defining the models. This is further
restricted when the possible variations in the spectral tilt
allowed by the COBE data are constrained by the higher $\ell$
Boomerang data. More constrictions arise from galaxy-galaxy and
cluster-cluster clustering observations: the shape of the linear
density power spectrum must match the shape reconstructed from the
data. In \cite{lange00,jaffe00,bmaxiboom00,boom01}, 
the LSS data were characterized by a simple shape
parameter and $\sigma_8$, with distributions broad enough 
so that these prior probability choices
would not be controversial, but encompass the ranges that almost
all cosmologists would believe.

Using all of the current CMB data and the LSS priors,
\cite{boom01} got $H_0=56\pm 9$, $\Omega_{\Lambda}=0.55\pm 0.09$,
with $n_s$, $\Omega_{tot}$ and $\Omega_bh^2$ virtually unchanged
from the CMB-only result. Apart from the significant
$\Omega_{\Lambda}$ detection, the dark matter is also strongly
constrained, $\Omega_{cdm}h^2 = 0.13^{+.03}_{-.02}$.
Restricting $\Omega_{tot}$ to be unity as in the usual inflation
models, with CMB+LSS, $\Omega_b h^2= 0.021\pm 0.003$,
$\Omega_{cdm}h^2 = 0.13 \pm 0.01$, $\Omega_{\Lambda}=0.62\pm 0.07$
and $n_s$ = $0.98^{+.10}_{-.07}$. If the equation of state of the
dark energy is allowed to vary, $w_Q <-0.3 $ is obtained, becoming
substantially more restrictive when supernova information is
folded in, $w_Q < -0.7$ at 2 $\sigma$ \cite{bmaxiboom00}.

\begin{figure}
\centerline{\epsfxsize=5.5in\epsfbox{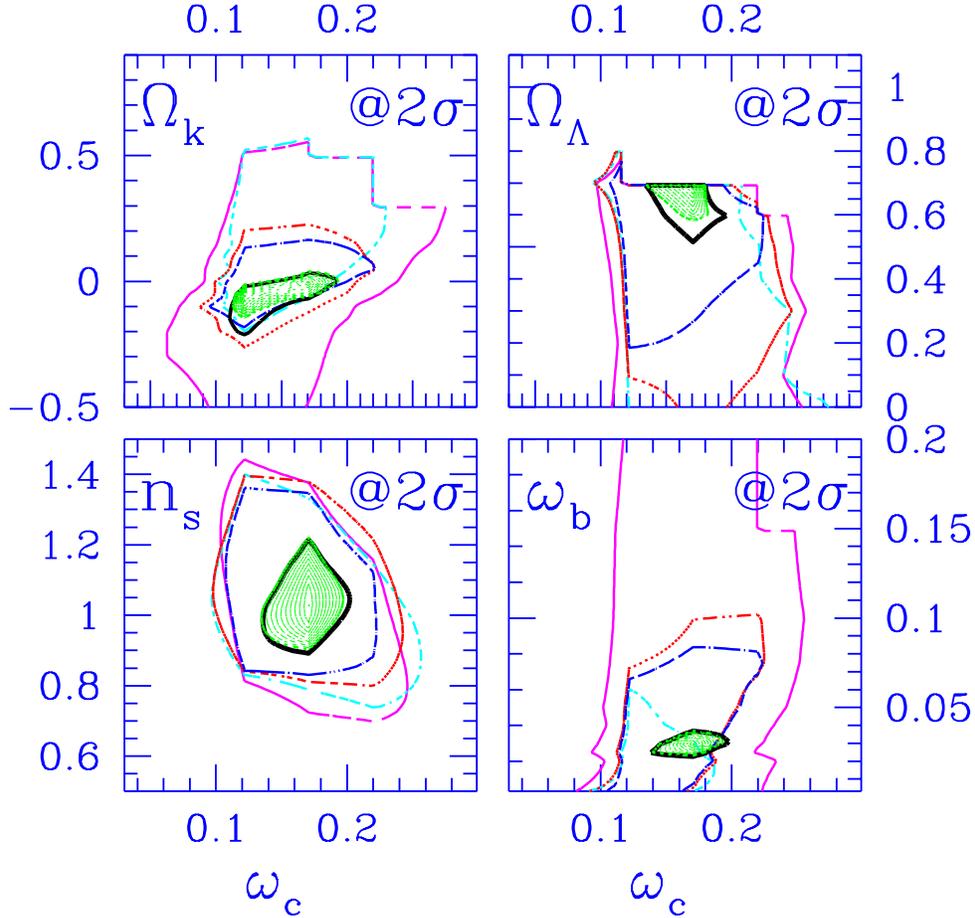}}
\vspace{-10pt} \caption{ This figure from
{\protect\cite{bmaxiboom00}} shows 2-$\sigma$ likelihood contours for
the dark matter density $\omega_c =\Omega_{cdm}{\rm h}^2$ and
$\{\Omega_k,\Omega_\Lambda,n_s,\omega_b\}$ for the LSS prior
combined with weak limits on $H_0$ (45-90) and cosmological age
($> 10$ Gyr), and the following CMB experimental combinations: DMR
(short-dash); the ``April 99" data (short-dash long-dash);
TOCO+4.99 data (dot short-dash); Boomerang-NA+TOCO+4.99 data (dot
long-dash, termed ``prior-CMB"); Boomerang-LDB + Maxima-1 +
Boomerang-NA + TOCO + 4.99 data (heavy solid, all-CMB). These
$2\sigma$ lines tend to go from outside to inside as more CMB
experiments are added. The smallest 2-$\sigma$ region (dotted and
interior) shows SN1+LSS+all-CMB, when SNI data is added. For the
$\Omega_\Lambda$, $n_s$ and $\omega_b$ plots, $\Omega_{tot}$=1 has
been assumed, but the values do not change that much if
$\Omega_{tot}$ floats. The main movement with the most recent data
{\protect\cite{boom01}} is that $\omega_c$ localizes more around
0.13 in all panels, and the $\omega_b$ contour in the lower right
panel migrates downward a bit to be in better agreement with the
Big Bang nucleosynthesis result. } \label{fig:cmbLSS2sig}
\end{figure}

Fig.~\ref{fig:cmbLSS2sig} gives a visual perspective from
\cite{bmaxiboom00} on how the parameter estimations for the
adiabatic models evolved as more CMB data were added (but with
always the LSS prior included for this figure). With just the
COBE-DMR+LSS data, the 2-$\sigma$ contours were already localized in
$\Omega_{cdm}h^2$. Without LSS, it took the addition of Maxima-1
before it began to localize. $\Omega_k$ localized near zero when
TOCO was added to the April 99 data, more so when Boomerang-NA was
added, and much more so when Boomerang-LDB and Maxima-1 were
added. Some $n_s$ localization occurred with just ``prior-CMB"
data. $\Omega_bh^2$ really focussed in with Boomerang-LDB and
Maxima-1, as did $\Omega_\Lambda$. If only DMR plus the most
recent Boomerang results~\cite{boom01} to $\ell =600$ are used
(the limit in \cite{debernardis00}), the plot is rather similar.
However, when all recent Boomerang~\cite{boom01} results to $\ell
=1000$ are used, the inner contour in the $\Omega_{cdm}h^2$
direction sharpens up in all panels, and the $\Omega_{b}h^2$
contour lowers to be in the good agreement with the big bang
nucleosynthesis result indicated above.

\section{Computing Non-Gaussian Secondary Signals: the SZ Example}

 The  secondary fluctuations involve nonlinear processes,
and the full panoply of $N$-body and gas-dynamical cosmological
simulation techniques are being used to study them. 
It is realistic to hope that the thermal and kinetic
Sunyaev-Zeldovich effects can be understood statistically this way
with sufficient accuracy for CMB analysis, and this is the main
example used in this section. On the other hand, star-bursting
galaxies will be quite difficult to understand from simulations
alone, but CMB analysis will be greatly aided by their
point-source character for most experimental resolutions. Galactic
foregrounds, however, cannot be ``solved" by hydro calculations.

\begin{figure}
\centerline{\hspace{-0.1in}\epsfxsize=5.2in\epsfbox{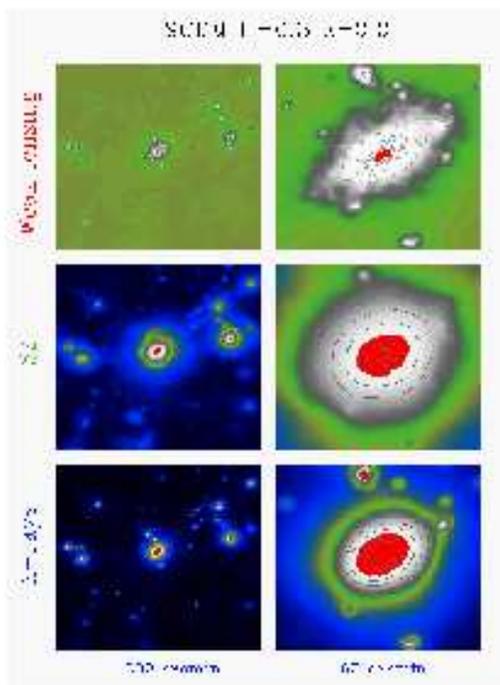} }
\caption{\baselineskip=10pt\small SZ maps for the ITP comparison
cluster seen at $z=0$ subtending the angles shown. The observing
wavelength was taken to be in the Rayleigh-Jeans region of the
spectrum, so $\Delta T/T =-2y$ here.  The core dark regions interior
to the white areas are above $32\times 10^{-6}$; the dark contours
surrounding the white are at $2\times 10^{-6}$, levels now accessible
to ground-based instrumentation. At higher redshift, gaseous filaments
bridge the subclusters which merge to make the final state. Even the 
far-field outskirts of these subclusters is observable, but precision below
$10^{-6}$ would be needed to probe well the
filaments. (See also Fig.~\ref{fig:SZpkscl}.)  }
\label{fig:SZitpclz0}
\end{figure}

\vs\noindent {\bf 2.1 Hydrodynamical Calculations}: The Santa
Barbara test \cite{itp95cl} compared simulations of an individual
cluster of $10^{15} \msun$ done with a variety of hydrodynamical
and N-body techniques, in particular grid-based Eulerian methods
and both grid-based and smooth particle (SPH) Lagrangian methods.
(Eulerian grids are fixed in comoving space, Lagrangian grids
adapt to the density of the flow.) Fig~\ref{fig:SZitpclz0} shows
this cluster, computed with the treePM-SPH code of Wadsley and
Bond \cite{wb01,bpkw98}, seen at the present in weak lensing, SZ
and X-rays, at contour thresholds designed to represent potentially 
observable levels. For a review of the great strides made in SZ
experimental methods and results on individual clusters up to 1998
see \cite{birkinshaw99}.

\begin{figure}
\vspace{-0.1in}
\centerline{\epsfxsize=5.0in\epsfbox{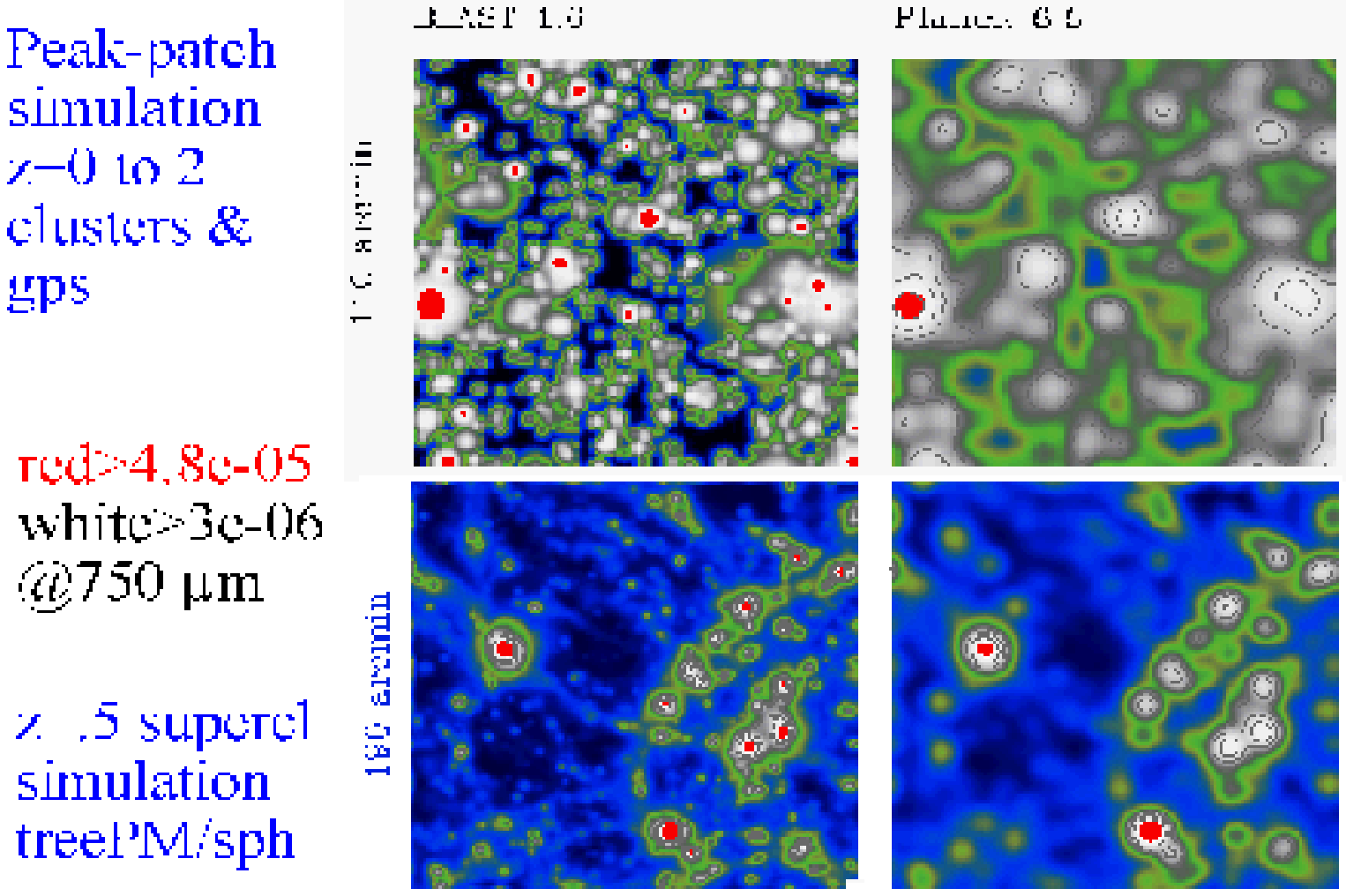} }
\vspace{-0.1in} \centerline{\hspace{0.1in}
\epsfxsize=5.2in\epsfbox{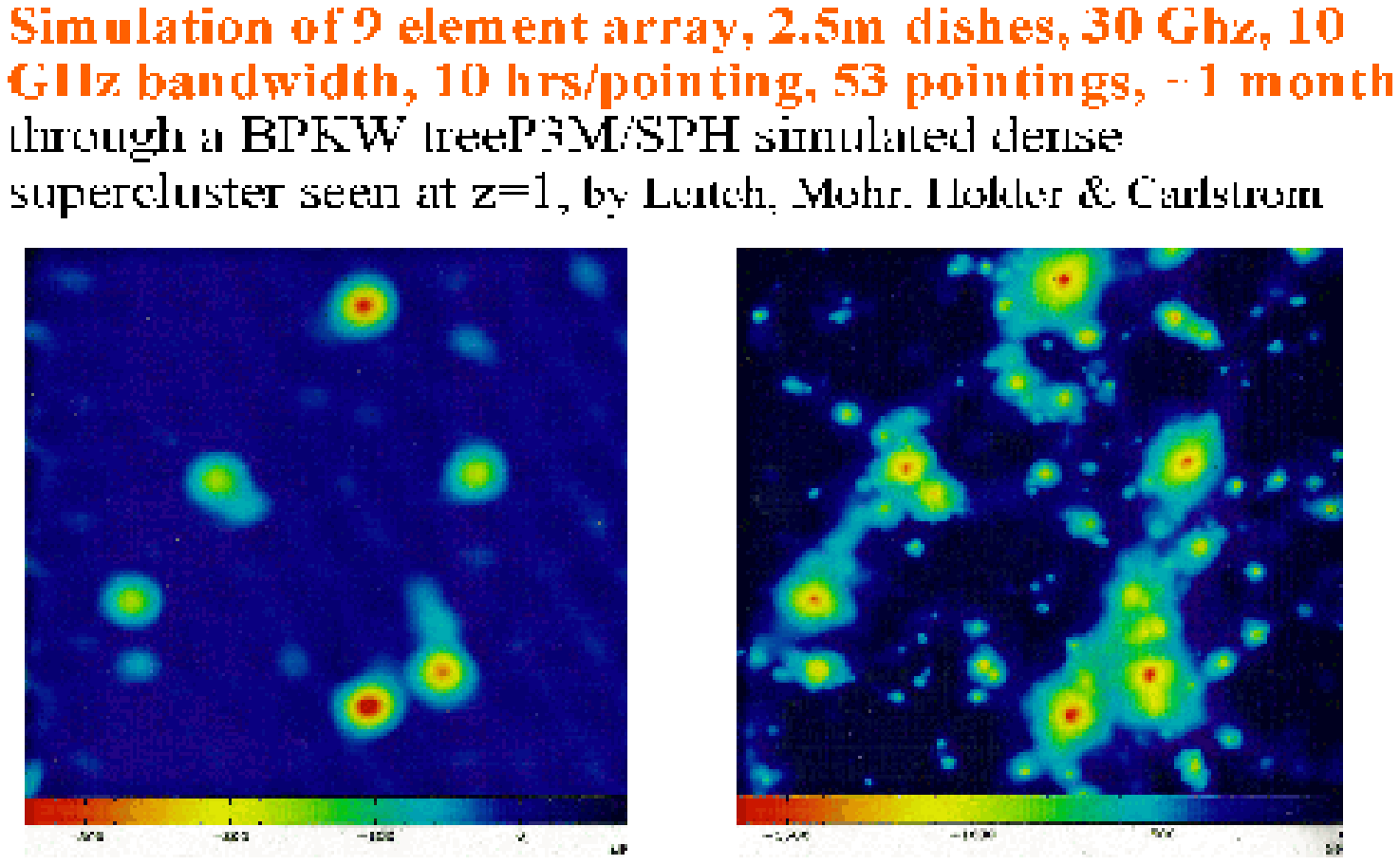} }
\vspace{-0.1in}\caption{\baselineskip=10pt\small SZ maps subtending
the angles shown for peak patches (above) and a single supercluster
region seen at redshift 0.5. The core dark regions interior to the
white areas are above $48\times 10^{-6}$, and the dark contours
surrounding the white are at $3\times 10^{-6}$ at 750$\mu m$, levels
that can be achieved with blank field SZ bolometer
experiments. Filamentary bridges between the clusters are typically
below $10^{-6}$, but the far-field of clusters can be probed. The
bottom left panel passes the supercluster region at $z=1$ in the right
panel through a simulated 30 GHz HEMT-based interferometer with the
indicated characteristics.}
\label{fig:SZpkscl}
\end{figure}

Another hydrodynamical example, the simulation of a supercluster
\cite{bpkw98,pbkw98} using the treePM-SPH code, is used to illustrate
the computational challenges once we extend beyond individual cluster
simulations. Fig.~\ref{fig:SZpkscl} shows thermal SZ maps of the
supercluster at redshift $z =0.5.$ The computation evolved from
redshift $z=30$ to the present a high resolution 104 Mpc diameter
patch with $100^3/2$ gas and $100^3/2$ dark matter particles,
surrounded by gas and dark particles with 8 times the mass to 166 Mpc,
in turn surrounded by ``tidal'' particles with 64 times the mass to
266 Mpc.

There were a total of 1.6 million particles including the
medium and low resolution regions, but much larger simulations are
possible in this era of massive parallelization. For example, a
parallel tree-SPH code (``gasoline''), with timesteps that can vary
from particle to particle, can routinely do $256^3$ gas and $256^3$
dark matter simulations, and even $512^3$ simulations on relatively
modest ($< 100 $ processor) BEOWULF systems~\cite{bwr01}.

The treePM technique is a fast, flexible method for solving
gravity and can accurately treat free boundary conditions. The SPH
part of the code includes photoionization as well as shock heating and
cooling with abundances in chemical (but not thermal) equilibrium,
incorporating all radiative and collisional processes. The species
considered were H, H$^{+}$, He, He$^{+}$, He$^{++}$ and e$^{-}$. The
extra cooling associated with heavy elements injected into the
intracluster media during galactic evolution was ignored, but so was
the feedback on the medium from energy injected from galactic
supernovae: the computation does not have high enough spatial
resolution to calculate galaxy collapse.

We now describe some aspects of simulation design for the supercluster
simulation \cite{bpkw98,pbkw98} which are also relevant for the
larger parallel calculations. One first decides on the mass resolution
to achieve, which is set by the lattice spacing of particles on an
initial high resolution grid, $a_L$, chosen to be $\sim 1 \mpc$
comoving to ensure that there will be adequate waves to form the
target objects, in this case clusters. \footnote{The mass resolution
limits the high $k$ power of the waves that can be laid down in the
initial conditions (Nyquist frequency, $\pi/a_L$), but for aperiodic
patches there is no constraint at the low $k$ end: the FFT was used
for high $k$, but a power law sampling for medium $k$, and a log $k$
sampling for low $k$, the latter two done using a slow Fourier
transform, \ie a direct sum over optimally-sampled $k$ values, with
the shift from one type of sampling to another determined by which
gives the minimum volume per mode in $k$-space. By contrast, more
standard periodic ``big box'' calculations are limited by the
fundamental mode, although one can alleviate this by nesting the high
resolution box in lower resolution ones as in the supercluster
simulation. Even so, this simulated supercluster could not be done in
a periodic code with this number of particles because of gross
mishandling of the significant large scale tidal forces.}  Next one
needs to determine the spatial resolution of the gas and of the
gravitational forces, preferably highly linked. In Lagrangian codes
like treePM-SPH, this varies considerably, being very high in cluster
cores, moderate in filaments, and not that good in voids. Since
accurate calculations of the cluster cores were a target, resolution
better than $\sim 40 \kpc$ was desired. The best resolution obtained
was about $15 \kpc$. Given $a_L$, the size of the high resolution part
of the simulation volume is determined by CPU limitations on the
number of particles that can be run in the desired time. This means
the high resolution volume may distort considerably during the
simulation. To combat this, progressively lower resolution layers were
added to ensure accurate large scale tides/shearing fields operated on
the high resolution patch.  For the calculations shown, the high
resolution region (grid spacing $a_L$, $100^3$ sphere) sits within a
medium resolution region ($2\,a_L$, $80^3$), and in turn within a low
resolution region ($4\,a_L$, $64^3$). The mean external tide on the
entire patch is linearly evolved and applied during the calculation.

\begin{figure}
\vspace{-0.2in}
\centerline{\hspace{0.1in}
\epsfxsize=5.0in\epsfbox{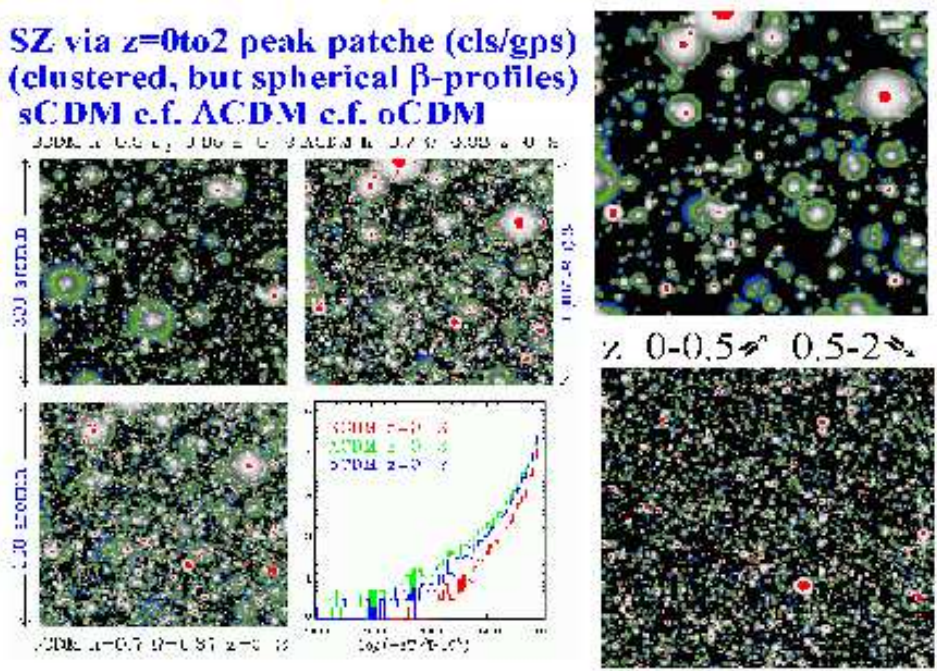} }
\vspace{-0.2in}
\caption{\baselineskip=10pt\small SZ maps derived from spherical
presure-profiles imposed upon halos identified up to $z$=2 with the
peak-patch method, for three cluster-normalized cosmological
models. The observing wavelength and contours are the same as in
Fig.~\ref{fig:SZitpclz0}. The histograms show $\Lambda$CDM, $o$CDM and
sCDM can be clearly differentiated. Blank field SZ surveys using
interferometers and bolometer experiments promise to revolutionize our
approach to the cluster system, especially at $z\gta 0.5$. The
contributions from clusters below and above this redshift are
shown on the right for the $\Lambda$CDM model. }
\label{fig:SZitpclz0-2}
\end{figure}

\vs\noindent {\bf 2.2 Non-Gaussian Source Model}: It is clear from
Figs.~\ref{fig:SZitpclz0-2} and \ref{fig:SZpkscl} that the dominant
signals are quite patchy if one is interested in amplitudes above a
few $\mu K$.  This has the disadvantage for analysis that non-Gaussian
aspects of the predicted patterns are fundamental, and an infinite
hierarchy of connected $N$-point spectra beyond the 2-point spectrum
$C_\ell$ are required to specify them (as for defect models). However,
it also seems reasonable that we could represent the results in terms
of extended emission from localized sources.  This concentration of
power is characteristic of many types of secondary signals. Indeed
some, such as radiation from dusty star-burst regions in galaxies, can
be treated as point sources since they are much smaller than the
observational resolution of most CMB experiments.  However, highly
accurate emission from dusty galaxies is too difficult to calculate
from first principles, and statistical models of their distribution
must be guided by observations.

Extended and point source models for the temperature fluctuations
express the signal as a projection along the line-of-sight of a 3D
random field, ${\cal G} ({\bf r}, t)$, which is the convolution of
profiles $g({\bf r} \vert {\cal C}, t)$ surrounding objects of some
class ${\cal C}$, with the comoving number density $n_{{\cal C}*}
({\bf r}, t)$ defining a point process for those
objects~\cite{bh95,b88}.
In \cite{b88,bch2}, the points
were referred to as ``shots", from shot-noise, although of course
continuous clustering of the shots as well as their uncorrelated
Poissonian self-correlation has to be included. Simple derivations for
the form of ${\cal G} ({\bf r}, t)$ and $g$ and the relation of the 2D
power spectrum ${\cal C}_\ell$ to the 3D power spectrum of the ${\cal
G}$ (and hence of $n_{{\cal C}*}$) are given in \cite{bh95}.

For dusty star-burst regions, the shots are galaxies and the profile
$g$ involves the dust density and temperature \cite{b88,bch2},
while for the SZ effect, the dominant emission comes from clusters and
groups and the profile $g$ is a constant times the pressure
\cite{b88,bh95,bm96}.

Various attempts have been made to use hydrodynamical codes to do 
an entire region up to $z=2$ or so.  However, it should be clear from
the discussion of the numerical setup that approximations are always
needed to do this. One approach was to calculate the full pressure
power spectrum from a single simulation, $\avrg{\vert \widetilde{{\cal
G}}(k,t)\vert^2}$, and appropriately project it to determine the SZ
${\cal C}_\ell$ \cite{persi95,refregier00}. A more ambitious approach
was to tile the space with boxes out to some redshift
\cite{scaramella93,pen00,dasilva00,springel00,bwr01}, but because of
computer limitations and resolution requirements, the boxes were
relatively small and computationally expensive. To make predictions,
tricks were required, \eg ``observing' translated and rotated versions
of the few  simulated boxes at many different output redshifts.

\vs\noindent {\bf 2.3 The Shots and their Profiles}: The shot
model provides a way forward semi-analytically, by laying down
pressure profiles around a catalogue of halos. The halos and their
properties could be computed with $N$-body-only simulations using
much larger boxes than those required for hydro. Another approach
is to use ``peak-patches'', a method for identifying halos in the
initial conditions of calculations which has been shown to be
accurate for clusters \cite{bm96}. For this approach, the entire
space can be tiled with contiguous boxes that are smoothly joined,
and have all the required long wavelength power to treat the
really rare halo concentrations, or ``super-duper-clusters'', that
can appear.

What to do for the pressure profiles, $p({\bf r} \vert {\cal C} , t)$, 
is of course debatable. One strategy is to use
profiles calculated from hydro simulations, but what has invariably
been done is even simpler: pressure profiles scaled by bulk halo
properties of the observed form derived from low redshift X-ray
cluster data (and extrapolated to the more poorly known high redshift
regime). For example, adopting a spherically symmetric isothermal
``beta'' profile is common: $p({\bf r} \vert {\cal C} , t) = p_c
(1+r^2/r_{core *}^2)^{-3\beta /2} \vartheta (R_{{\cal C}*} - r)$,
where $p_c$ is the central pressure, $r_{core*}$ is a core radius, and
$R_{{\cal C}*}$ is an outer truncation radius exterior to which the
local pressure contribution is taken to be zero, and interior to which
$\vartheta$ is unity. The radius $r$ here is comoving and the $*$
subscript denotes comoving radii. $\beta \approx 2/3$ is a reasonable
fit to the X-ray data.

This rapid semi-analytic peak-patch method for simulating clusters
using isothermal beta-profiles was used in the construction of the two
SZ layers (thermal and kinetic) of the sandwich in
Fig.~\ref{fig:spatialforegrounds}.  The panels of
Fig~\ref{fig:SZitpclz0-2} show peak-patch-derived SZ maps for three
cosmologies, and the lower right corner shows the distribution of $\Delta
T/T$ in pixels, with its non-Gaussian distinct tails. The right two
panels contrast the contribution for the $\Lambda$CDM example of
clusters below  $z = 0.5$ with those above. The top panel of
Fig.~\ref{fig:SZpkscl} shows the $\Lambda$CDM example with a filter
scale appropriate for the Planck satellite resolution and with an
ideal resolution for the cluster system, below two arcminutes, which
interferometers and large single dishes can achieve. However, interferometers
typically filter the low $\ell$, losing large scale power.
The lower panel of Fig.~\ref{fig:newtnatocl9} shows the SZ ${\cal
C}_\ell$ spectrum derived for the flat $\Lambda$CDM simulation shown,
contrasting the contributions from clusters and groups above $z = 0.5$
with all of them.  Both Poissonian and clustering contributions
are included, since the simulation has them.

The power is not distributed in a democratic fashion as
it is for Gaussian theories, but is concentrated in cold spots below
218 GHz, and in hot spots at higher frequencies. 
Thus, the naive visual comparison with
bandpowers derived using a Gaussian assumption about the distribution
of power is misleading.

The assumption that the SZ effect is dominated by the high
pressure regions associated with clusters and groups of galaxies, with
only weak modifications coming from the intercluster medium, is borne
out visually in the supercluster simulation. The emission from the gas
in filaments outside of the groups and poor clusters that reside in
them is weak, below observable levels. What we can also see, however, is
that an anisotropic pressure distribution is expected, with elongation
along the filament directions, inviting improvements over spherical
approximations.

To properly analyze the SZ effect in ambient fields, having fast Monte
Carlo simulation methods along the lines developed here is clearly
essential. That does not take away from the challenge
of how best to do the analysis of such distinctly non-Gaussian
signals, which consist of extended rather than point-like sources.
Even with relatively small beams, these will be somewhat confused by
overlapping sources. At least with the cluster system, we may hope
to model the individual elements with hydrodynamical
simulations. For other secondary sources such as dusty galaxies, a
priori theoretical models are not really feasible, and the properties
of the sources will have to be derived from the observations. At least
in some of those cases, for typical beam sizes, the emission can be treated as
from point sources.

\vs\noindent {\bf 2.4 Foreground Complications}: The {\it
foreground} signals from the interstellar medium are also
non-Gaussian, but are not well-modelled by extended sources, 
since their emission power spectrum ${\cal C}_\ell$ has
been shown to rise to lower $\ell$; further, no simplifications
such as statistical isotropy apply. Direct hydrodynamical studies
may increase our understanding of the ISM, but will not be able to
provide strong guidance on statistical comparisons with data. The
use of other data sets will clearly play a strong role in the
final analysis; \eg templates from the IRAS+DIRBE data is useful
for modelling the distribution of shorter wavelength dust
radiation. Exactly what to do for the analysis of foregrounds is
under intense study and some current ideas are described in
\S~\ref{sec:act}.

\section{Acting on the Data}\label{sec:act}

The data comes in as raw timestreams, which must be cleaned of
cosmic rays, obvious sectors of bad data, and other contaminants.
Encoded in it are the sky and noise, as well as unwanted instrument,
atmospheric and other residual signals. The pointing matrix identifies 
time-bits to a chosen pixel basis and allows the spatial signal to be
separated from the noise; the resulting map completely
represents the data if the pixelization is fine enough relative to
the experiment's beam. At one time, the cosmology was drawn
from the maps, perhaps using new bases beyond the spatial pixel
ones, and possibly truncated to {\it compress} the data into more
manageable chunks (\eg using signal-to-noise eigenmodes to get rid
of informationless noisy components of the map, \cite{b94}). 

The statistical distribution of various operators on the pixelized data
can be estimated from the map.  In particular, the power spectrum
${\cal C}_\ell$ represents a {\it radical compression} of
the data, but of course is no longer a complete representation of
the map. However, under the assumption that the sky maps only
contain Gaussian signals characterized by an isotropic power
spectrum, this is
a complete representation of the statistical information in the
map when the binning of the bandpowers is fine enough. 
For such Gaussian theories, such as those usually arising in
inflation, cosmological parameters can be quickly estimated
directly from the power spectrum. The radical compression has
allowed millions of theoretical models to be confronted with the
data in the large cosmological parameter spaces described above,
\eg \cite{lange00}, a situation infeasible if full statistical
confrontation of all the models with the maps was required.

Basic aspects of the Boomerang, Maxima, DASI and CBI pipelines are
described in \cite{bcjk99},
\cite{prunet00,dore01,boom01,master01},
\cite{maximapipeline00,maximapipeline01}, \cite{dasi01} and
\cite{cbi00}. Pipelines were also developed for DMR and for QMAP.
The MAP pipeline as currently envisaged is described in
\cite{hinshaw00}. The Planck pipeline is still very much under
discussion and development, complicated by the necessity of
delivering high precision component maps and derived parameters
from such a huge volume of data.

\vs \noindent {\bf 3.1 Bayesian Chains}: The data analysis
pipeline can be viewed as a Bayesian likelihood chain:
\begin{eqnarray}
&& \cP(parameters \vert data, \ theoretical \ framework) \nonumber\\
&& = \cP(data \ timestreams \vert maps,  \ noise) \nonumber\\
&& \ \ \otimes \cP(noise\vert
noise-obs) \nonumber \\
&&\quad \otimes  \cP(maps \vert signals) \nonumber\\
&& \qquad \otimes \cP(signals \vert parameters) \nonumber\\
&& \ \qquad \otimes \cP(parameters\vert prior \ knowledge)
\nonumber
\\
&&\ \ \qquad \otimes [1/\cP(data \vert theoretical \ framework)]
\end{eqnarray}
Each conditional probability in the unravelling of the chain is
very complex in full generality. This means the exact statistical
problem is probably not solvable, and only applying simplifying
approximations to sections of the chain have made it tractable. Of
particular widespread use has been the assumption that both noise
and signals are Gaussian-distributed; others are being
explored, especially by Monte Carlo means, but much remains to be
done for us to attain the high precision levels that have been
forecast.

We now describe the various terms, beginning with the last two. The
priors $\cP(parameters\vert prior \ knowledge)$ for the cosmological
parameters are often taken to be uniform, but can represent
information from other experiments, \eg LSS observations, prior CMB
experiments, age or $H_0$ constraints.  The hope is that the new
information from the rest of the probability chain will be so
concentrated that the precise nature of the prior choice will not
matter very much. As we have seen, prior information from non-CMB
experiments is crucial, even given the most ideal CMB experiment, in
order to break near-degeneracies among cosmological parameters.

The quantity $\cP(data \vert theoretical \ framework)$ is an
overall normalization. It can address whether the overall theory
is crazy as an explanation of the data. It has some
characteristics in common with goodness-of-fit concepts in
``frequentist" probability analysis.

We often refer to entropies  ${\cal S} \equiv \ln \cP$ in the
Gibbs sense so the Bayesian chain involves a sum of the individual
conditional entropies that make up the whole.

\vs \noindent {\bf 3.2 Signal/noise separation}:
The information comes in the form of a discretized timestream, with
$d_{c t}$ as the data for the bit of
time $t$ in frequency channel $c$, and a pointing vector on the sky,
$\hq_c (t)$. 
The sky is gridded into pixels, and these are used to define signal maps 
which are functions of direction and frequency.
We denote these by $\Delta_{c p}$, for the signal at pixel $p$ and frequency
channel $c$. The pointing matrix, $P_{c tp}$, is an operator mapping
time-bit $t$ to pixel $p$ with some weight. The difference $d-P\Delta$
is the noise, $\eta$, plus further residuals ${\cal R}$, which may be
sky-based or experimental systematics. The probability of a
given timestream can be written as
\begin{equation}
  \label{eq:data}
 \cP(d \vert
\Delta, \eta, P_{ctp}) = \prod_{ct} \delta \big( d_{c t} -
 ( \sum_p P_{ctp} \Delta_{c p} +{\cal R}_{ct} + \eta_{c t} )\big) .
\end{equation}

\noindent {\bf 3.2.1 Map-making}: Map-making for a given channel
involves using just the information encoded in $d$ and $P$ to
separate what is sky signal (and residuals) from what is noise,
\ie the construction of $\cP (\Delta \vert d)$.

The pointing matrix is a projection operator, with many more time-bits
 than pixels. The simplest example for $P_{c tp}$ is the
characteristic set function for the pixel $\chi_p (\hq_c (t))$ (one
inside the pixel, zero outside). For a chopping experiment such as
MAP, the two pixels where the two beams are pointing (separated by
$141^\circ$) are coupled at a given time-bit.  There are also choices
to be made about scattering the information in the timestream among
pixels. For example, one can include the beam in $\Delta$ or in
$P$. At the moment the former is preferred, but there are good
arguments for the latter if the beam changes in time (\eg asymmetric
beams with rotations). In that case, $\chi_p (\hq_c (t))$ is replaced
by $A_p {\cal B}(\hq_p - \hq_c (t))$, where ${\cal B}$ is the beam
function and $A_p$ is the pixel area, and the derived signal
$\Delta_{cp}$ does not have the beam function convolved with it as it
does in the ``top hat" $\chi_p$ case. Of course, any other function
normalized to the pixel area upon integration can be used, \eg
approximations to the beam which might not contain all sidelobe
information. As long as the signals are appropriately
convolved, and the pixelization choices are fine enough, the results
should be identical. Pointing uncertainty further complicates the
simplicity of $P$.

 In map-making, there is an implicit assumption on the prior
probability of the map, $\cP (\Delta \vert prior)$, that it is uniform
so any value is {\it a priori} possible. The map distribution should
involve integration over all possible noise distributions; in
practice, only the measured noise is used, which delivers the maximum
likelihood map $\overline{\Delta}_{cp}$ and a pixel noise correlation
$C_{N, cp c^\prime p^\prime}$, allowing for possible channel-channel
correlations in the noise.

For many previous experiments, one could just compute the average
$\overline{\Delta}$ and variance $C_{N}$ of the measured $d$'s
contribution to each pixel, relying on the central limit theorem
to ensure a Gaussian-distribution. However, these ``naive maps'',
described below in more detail,  also relied on an implicit
assumption, namely that $\eta$ was uncorrelated from time-bit to
time-bit, which is not correct with the $1/f$ noise in detectors.
The noise was also often so strong compared with the signal that
the latter could be ignored in solving for $\Delta_{cp}$; this is
no longer the case.

So far it has been essentially universal to assume that $\eta$ is
Gaussian-distributed stationary noise with noise power $w^{-1}(\omega )$
a smooth function. Thus, the noise is completely specified by its
covariance $c_{n,ct,c^\prime t^\prime} \equiv {\cal S}[\langle
\eta_{ct} \eta_{c^\prime t^\prime} \rangle ]$, which is
assumed to be a function of $t-t^\prime$, hence diagonal in the time
frequencies $\omega$. Here, ${\cal S}$ denotes the smoothing
operation. Drifts and non-Gaussian elements can be included with
timestream filtering and in the catch-all residual ${\cal R}$ using
time templates (see below). Ignoring ${\cal R}$ for the moment, the
probability of the time-ordered data given a map and noise-power $w^{-1}$,
${\cal P}(d|\Delta, w)$, is just a Gaussian:
\begin{eqnarray}  \cP (d\vert \Delta , w)
&=&\exp[-(d-P\Delta)^\dagger w (d-P\Delta)/2]/\{(2\pi
)^{N_{tbits}/2}
\sqrt{{\rm det}(w^{-1})}\} \nonumber \\
&=&\exp[-(\overline{\Delta}-\Delta)^\dagger W_N (
\overline{\Delta}-\Delta)/2]/\{(2\pi )^{N_{pix}/2} \sqrt{{\rm
det}(W_N^{-1})}\} \nonumber \\
&&  \times \exp[-\bar{\eta}^\dagger
\widetilde{w}\bar{\eta}/2]/\{(2\pi )^{(N_{tbits}-N_{pix})/2}
\sqrt{{\rm det}(\widetilde{w}^{-1})}\} , \label{eqn:mapprobe} 
\end{eqnarray} 
\vspace{-0.25in} 
\begin{eqnarray} 
 W_N\overline{\Delta} = P^\dagger w d,\qquad \qquad \, \, W_N\equiv P^\dagger wP
\equiv C_N^{-1}= \avrg{n n^\dagger }^{-1} , \label{eqn:mapsoln}\\
 n \equiv \overline{\Delta}-\Delta =C_N P^\dagger w\eta ,  \qquad \qquad \qquad 
 \bar{\eta} \equiv \avrg{\eta \vert d} = d-P\overline{\Delta}, \nonumber \\
 \widetilde{w} = w(1-P(P^\dagger w P)^{-1}P^\dagger w), \qquad \qquad \qquad  
\qquad P^\dagger w \bar{\eta}=0 . 
\label{eq:meannoise}
\end{eqnarray}
The maximum likelihood solution for $w(\omega )$ is $({\cal S}[\vert
{\rm FT}(d-P\overline{\Delta})\vert^2])^{-1}$, where FT denotes
the Fourier transform operator. Note that the $\bar{\eta}$ noise
term is multiplied by a projector which is perpendicular to the
spatial sector; \ie $C_NP^\dagger w \bar{\eta}$=0. The first
Gaussian in eq.(\ref{eqn:mapprobe}) is $\cP (\overline{\Delta}
\vert \Delta)$. We can interpret it as the integration over the
spatial Gaussian random noise field $n$ defined in
eq.(\ref{eqn:mapsoln}) of the product of $\cP
(\overline{\Delta} \vert \Delta , n)= \delta (\overline{\Delta} -(
\Delta +n))$ and the distribution of $n$, $\exp[-n^\dagger C_N^{-1}n/2] /
[(2\pi)^{N_{pix}/2}[{\rm det} C_N]^{1/2}]$ .

\noindent {\bf 3.2.2 General {\it cf.} Optimal Maps}: Any linear
operation from timestream to generalized pixels,
$\overline{\Delta}^{(\cM)} = \cM^\dagger d$ defines a map, albeit
not an optimal one as for the maximum likelihood solution
eq.(\ref{eqn:mapsoln}). The map-noise $n^{(\cM)} \equiv
\overline{\Delta}^{(M)} - {\Delta}^{(M)}$ has a correlation matrix
$C^{(\cM)}_N=\cM^\dagger w^{-1} \cM$. Here the relation of the
true sky signal seen in the $\cM$-filter-space to the ``optimal
map" signal $\Delta$ is ${\Delta}^{(M)} = \cM^\dagger P \Delta $.
Of special interest are therefore classes of operators $\cM$ for
which $\cM^\dagger P = P^\dagger \cM $ is a projector, \ie
self-adjoint and equal to its square. For example, $\cM^\dagger =
(P^\dagger w_* P)^{-1} P^\dagger w_* $ satisfies $\cM^\dagger P =
I$, the identity, for any $w_*$ provided $(P^\dagger w_* P)$ is
invertible. The price one pays for such a non-optimal map is
enhanced noise, with noise-weight matrix $W_N^{(\cM)} =(P^\dagger
w_* P) (P^\dagger w_*w^{-1}w_* P)^{-1} (P^\dagger w_* P)$ which
has maximum eigenvalues if $w_*=w$, but is otherwise ``less
weighty''.

An interesting example is when $w_*$ is taken to be a constant and
$P_{tp}$ is zero or one depending if the pointing at a time-bit
$t$ lies within the pixel $p$ or not; in that case $P^\dagger P$
just counts the number of time-bits that fall into each pixel, and
$\overline{\Delta}^{(\cM)}$ is just the average of the
anisotropies over these. If we use $\eta_t \eta_{t^\prime}$ in
place of its smoothed version $w^{-1}$, the noise matrix just
counts the variance in the amplitudes of the pixel hits about the
mean. For this reason, $(P^\dagger P)^{-1}P^\dagger d$ is called a
``naive map'' \cite{boom01}.  For other cases, computing the error
matrix may not be simple if $w$ is not known.

For the full maximum likelihood map, even with stationarity, the
convolution of $w$ with $d$ appears to be an $O(N_{tbits}^2)$
operation, but it is really $O(N_{tbits})$ because $w(t-t^\prime )$
generally goes nearly to zero for $t-t^\prime \gg 0$. Similarly,
the multiplication of the pointing matrix is also $O(N_{tbits})$
because of its sparseness. Thus, we can reduce the timestream data
to an estimate of the map and its weight matrix in only
$O(N_{pix}^2)$ operations.

\noindent {\bf 3.2.3 Iterative Map-making Solutions}:  An
iterative method \cite{prunet00} has proved quite effective to
solve this: the map on iteration $j+1$ is estimated from $W_*
\Delta_{(j+1)}=W_* \Delta_{(j)} +P^\dagger w_{(j)} \eta_{(j)}$,
where the noise and noise power on iteration $j$ are determined
from $\eta_{(j)} = d-P\Delta_{(j)}$, $w_{(j)}^{-1}(\omega) = {\cal
S}[\vert FT(\eta_{(j)})\vert^2 (\omega)]$. Here $W_* \equiv
P^\dagger w_* P$ is a matrix chosen to be easy to invert: \eg a
constant $w_*$ was used for Boomerang \cite{prunet00}. As the maps
converge, the noise orthogonality condition holds, so the solution
is the correct one. What $w(\omega)$ looks like is white noise at
high temporal $\omega$, crashing to zero because of $1/f$ noise in
the instruments and time-filtering. These low frequencies
correspond to large spatial scales, and the iteration converges
slowly there, inviting multigrid speedup of the algorithm, which
is now being implemented \cite{dore01}. The final
$\overline{\Delta}$ and $C_N$ are computed using $d$ and the
converged $w_{(j)}$ in eq.(\ref{eqn:mapsoln}). In some instances,
we can work directly with $W_N\overline{\Delta}$ and $W_N$ rather
than $\overline{\Delta}$ and $C_N$, avoiding the costly
$O(N_{pix}^3)$ inversion. If sections of the map are cut out after
$W_N$ determination, through a projector $\chi_{cut}$ which is
zero outside and one inside the cut, the cut weight matrix is
$(\chi_{cut} W_N^{-1}\chi_{cut})^{-1}$. This corresponds to
integration over all possible values of the now unobserved
$\overline{\Delta}$'s outside of the cut, which increases the
errors on the regions inside because of the correlation.
Unfortunately, two matrix inversions are required, and the first
one is potentially quite large even if the cut is severe.

Hinshaw \etal \cite{hinshaw00} describe the current pipeline for
MAP. To a good approximation the MAP noise is white, so $w_{(j)}=w_*$
can drop out, and the critical element is computation of $P^\dagger
P$, which is more complex than counting pixel-hits because MAP is a
difference experiment. The MAP team choose $W_*$ to be $diag(P^\dagger
w_* P)$, but otherwise use the same iterative algorithm.

Since $P$ is all that differentiates signal from noise in the
timestream, it is essential that it be sufficiently complex; in
practice this means many cross linkages among pixels in the scan
strategy. This is because there are random long term drifts in the
instruments that make it hard to measure the absolute value of
temperature on a pixel, though temperature differences along the
path of the beam can be measured quite well because the drifts are
small on short time scales. If there are not sufficient
cross-links, the maps are {\em striped} along the directions of
the scan pattern, and the weight matrix $W_N$ has to be called
upon to demonstrate that these features are not part of the true
sky signal we are searching for. The rapid on-board differencing
for MAP effectively eliminates this.

\vs \noindent {\bf 3.3 Pixelization}: Pixels are any discrete
basis that give a complete representation of the data. Square
top-hat functions tiling the space in a grid quite a bit smaller
than the beam size were the norm in the past. As the size of the
data increased, more elaborate hierarchical schemes were needed.
For COBE, the Quadrilateralized Spherical Cube was used: the sky
was broken into six base pixels corresponding to faces of a cube.
Higher resolution pixels were created by dividing each pixel into
four smaller pixels of approximately equal area, in a tree with
pixels that are physically close having their data stored close to
each other. There are (quite) small errors associated with the
projections of the sphere onto the faces. At resolution $r$, there
are $6 \times 2^{2(r-1)}$ pixels. The COBE data came with
resolution $r=6$, 6144 pixels of size $2.6^\circ$, and we
sometimes use resolution 5, still safely smaller than the
$7^\circ$ beam.

For large sky coverage, it is beneficial to have a pixelization
which is azimuthal, with many pixels sharing a common latitude, to
facilitate fast spherical harmonic transforms (FSHTs) \cite{mnv}
between the pixel space, where the data and inverse noise matrix
are simply defined, and multipole space, where the theories are
often simple to describe. The FSHT operation count is a ${\cal O}(
N_{pix}^{3/2})$ operations, with ${\cal O}( N_{pix}\ln N_{pix}^2)$
potentially possible to achieve. HEALPix\cite{ghw} is an example
which has been adopted for Boomerang, MAP and Planck: it has a
rhombic dodecahedron as its fundamental base, which can be divided
hierarchically while remaining azimuthal. There is also an
extensive software package that accompanies it to allow
manipulations. See also \cite{igloo} for the alternative
hierarchical ``igloo'' choice, with many HEALPix features, and
some other advantages.

In some cases {\it generalized pixels} are used, involving direct
projection of the data onto spatially-extended mode functions. The
first extensive use of this was for the SK95 dataset.
Signal-to-noise eigenfunctions used in data compression for COBE
and SK95 are another example \cite{b94,bjk98}. Direct projection
onto pixels in wavenumber space, with the mode functions being
discrete Fourier transform waves, also form a fine alternative
basis. Still there is nothing like a real-space visual
representation to explore strange features in the data.

\vs \noindent {\bf 3.4 Filtering in Time and Space}: Getting rid
of unwanted systematic effects in the data has been fundamental
throughout the history of CMB experiments. Sometimes this was done
in hardware, \eg rapid Dicke switching from one direction in the
sky to another in ``two-beam" chop experiments and ``three-beam"
chop-and-wobble experiments; sometimes it was done in editing, \eg
bad atmospheric contamination, cosmic rays; and now it is
invariably done in software, \eg $1/f$ noise in bolometers or
HEMTs, spurious spin-synchronous signals associated with rotating
platforms. In the linear operation $\sum_{t^\prime} {\cal
F}_{tt^\prime} d_{t^\prime}$ on the timestream, the filter matrix
${\cal F}_{tt^\prime}$ is often time-translation invariant, hence
diagonal in frequency. High pass filters are an example,
translating through the time-space pointing matrix $P_{tp}$ into
primarily low $\ell$ spatial filtering, with the penalty the
removal of part of the target sky signals as well as the unwanted
ones.

\noindent {\bf 3.4.1 Spatial Filters and Wavenumber Space}: In
direct filtering of the maps, the signal amplitude $ \Delta_p$ in
a pixel $p$ can be written in terms of generalized pixel-mode
functions ${\cal F}_{p,\ell m }$ and the spherical harmonic
components of the anisotropy, $a_{\ell m}$: $ \Delta_p =\sum_{\ell
m} {\cal F}_{p,\ell m } a_{\ell m}$. ${\cal F}_{p,\ell m }$
encodes the basic spatial dependence (\eg $Y_{\ell m} (\hq_p )$)
and filters, including the experimental beam ${\cal B}_{\ell m}$
and pixelization effects at high $\ell$, and the switching
strategy at low $\ell$.

\begin{figure}
\vspace{-0.2in}
\centerline{\hspace{0.00in}
\epsfxsize=2.95in\epsfbox{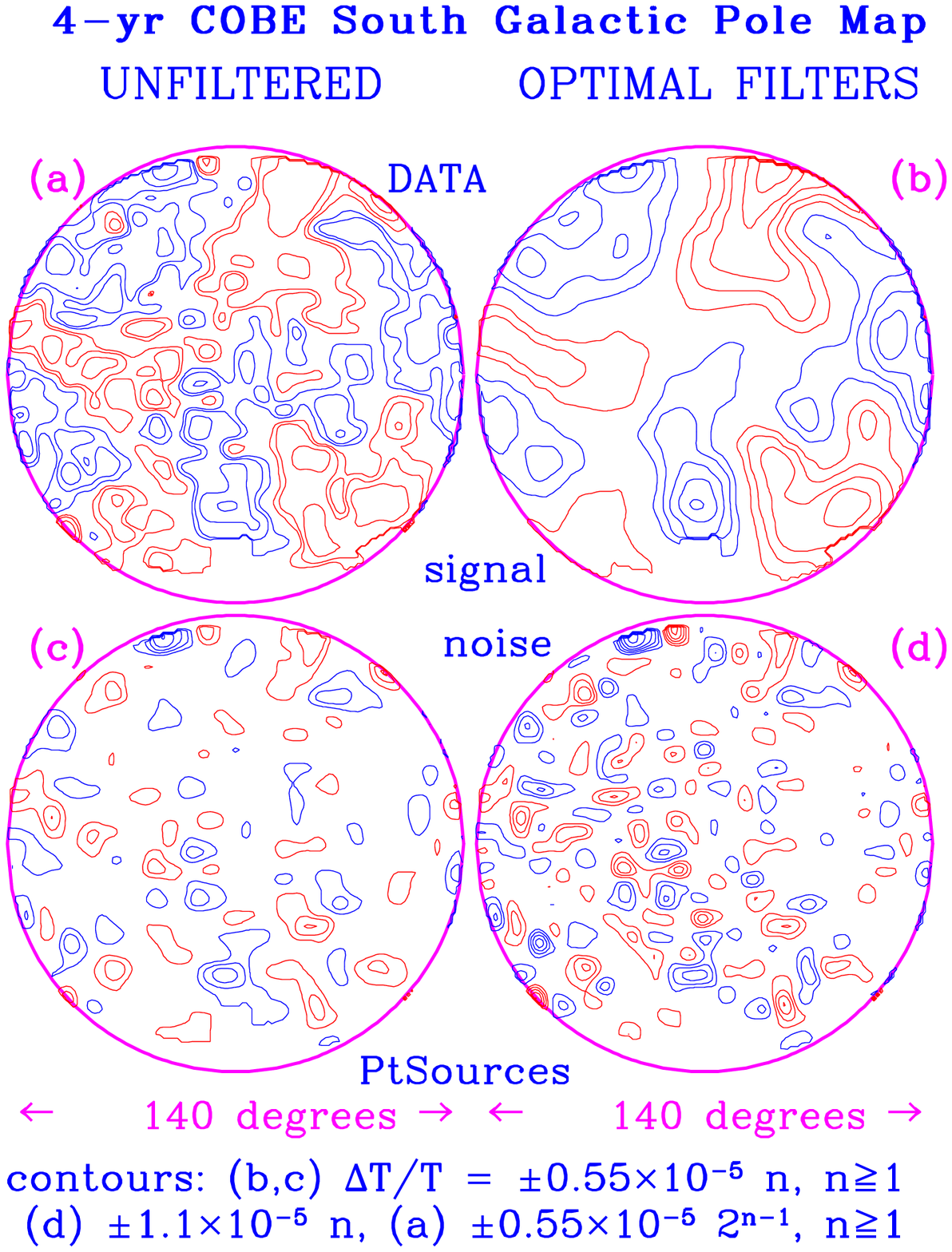}
\hspace{-0.45in}
\epsfxsize=2.95in\epsfbox{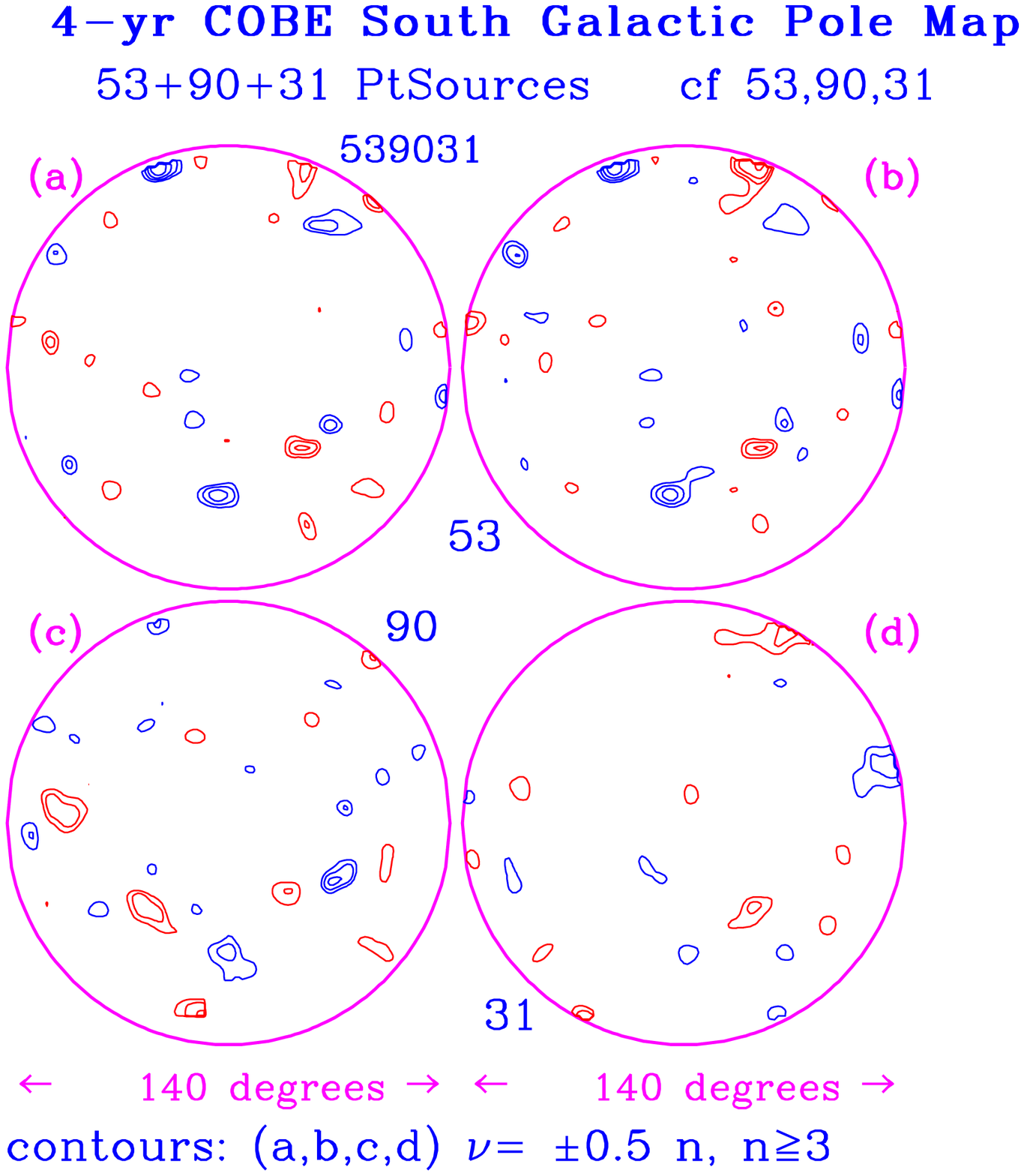} }
\vspace{-0.2in} \centerline{\hspace{0.00in}
\epsfxsize=2.95in\epsfbox{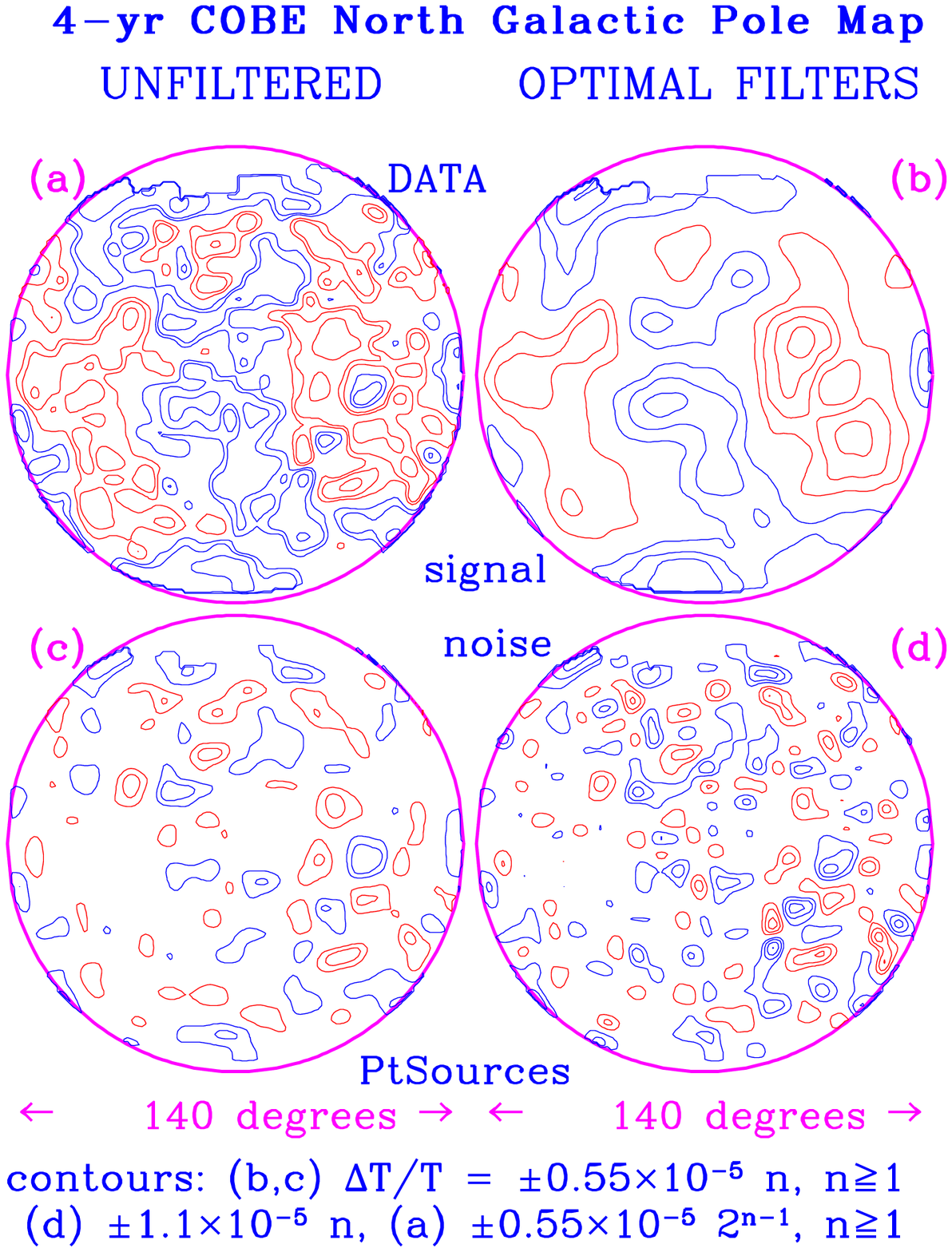}
\hspace{-0.45in}
\epsfxsize=2.95in\epsfbox{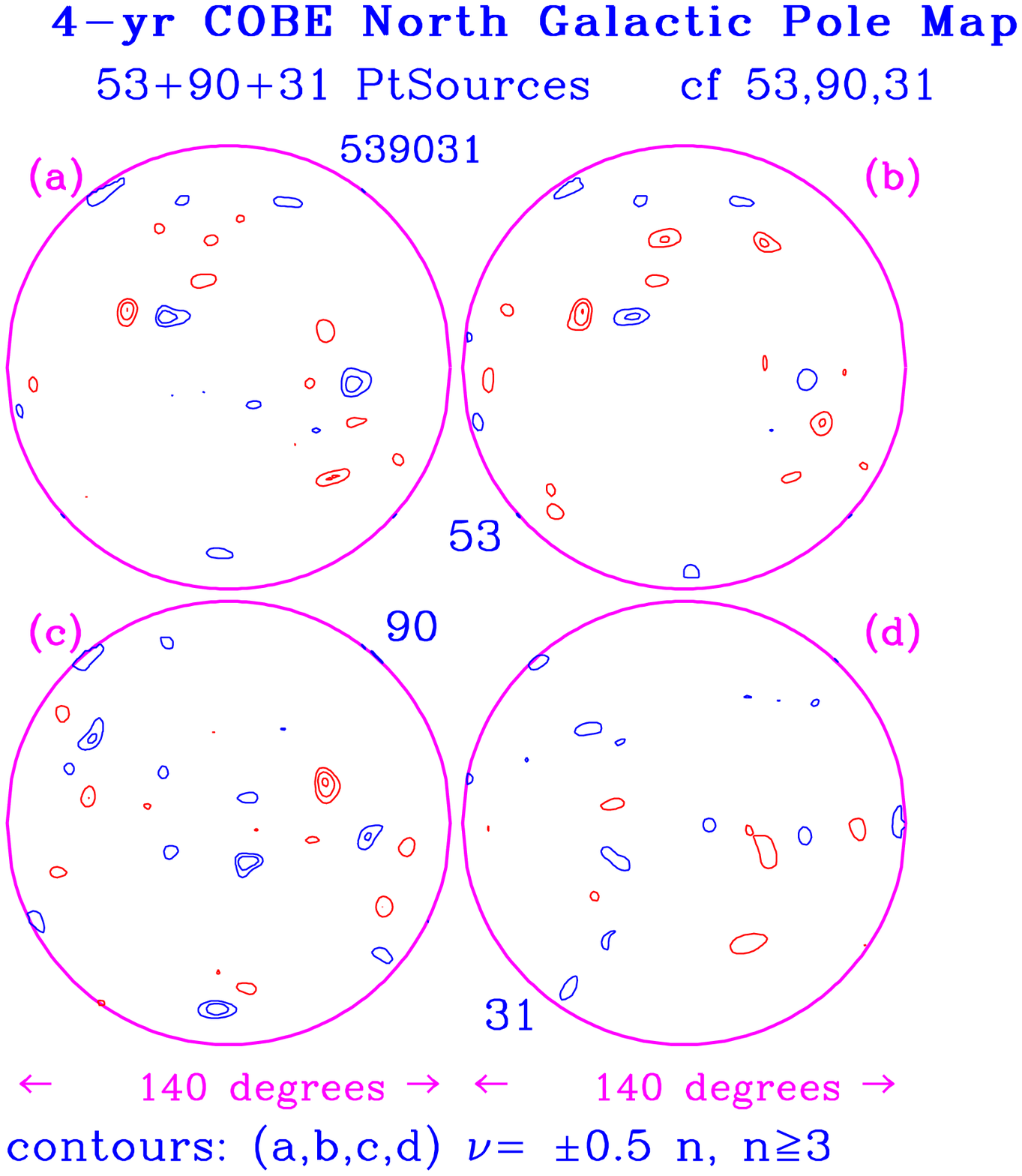} }
\vspace{-0.2in} \caption{\baselineskip=10pt\small The upper left panel
shows the unfiltered DMR map (combined 53+90+31 GHz data)
of a $140^\circ$ diameter region centered
on the South Galactic Pole, with the contours indicated, (b) and
(d) show the Wiener-filtered signal and noise maps, while (c) shows
the fluctuations of height $\nu \sigma$, with $\vert \nu \vert >
1.5$. The lower left panel shows the same for the North
Galactic Pole.  Note the similarity between (c) and (d), that is that
the Point Sources found at these low contours are consistent with
random concentrations of noise, the logical conclusion within the
statistics: there are no rare events that are good source candidates.
We chose such a low cut so
that one could see the effects of making $2$ or $ 3 \sigma$ cuts
just by counting in, each positive contour being 0.5 higher than
the previous. The right panels show the contours of $\nu$ rather than
$\Delta T/T$ for the individual (A+B) frequency maps. A true source
would be expected to persist in the three maps, as well as being in the
combined one, the strength variation across channels
depending upon the emission mechanism. Again there are no obvious
hits. Apart from being a useful way to detect point sources, this is a
nice mechanism for looking for anomalous local non-Gaussian
patterns, to be investigated to determine if we are dealing with
funny noise or true signals by appropriate follow-up investigations.
}\label{fig:SPsig}
\end{figure}

A Fourier transform representation of the signals in terms of
wavenumber ${\bf Q}$ is often justified. There is a projection from
the sphere with coordinates $(\theta , \phi)$ onto a disk with radial
coordinate $\pomega = 2 \sin (\theta /2)$ and azimuthal angle $\phi$
which is an area-preserving map and one-to-one --- except that one
pole gets mapped into the outer circumference of the disk. It is
highly distorted for angles beyond $180^\circ$, but even for the
$140^\circ$-diameter COBE NGP and SGP cap maps of Fig.~\ref{fig:SPsig}
it is an excellent representation. The associated discrete Fourier
basis for the disk is close to a continuum Fourier basis ${\bf Q}$,
with $\vert {\bf Q} \vert = \ell +1/2$.

Instead of a cap, consider a rectangular patch of size $L \times
L$. The amplitude $\Delta_p =\sum_{{\bf Q}} \widetilde{{\cal
F}}_{p} ({\bf Q})a_{\bf Q}$ then involves a mode-sum with
differential element $L^2 d^2{\bf Q}/(2\pi )^2$ = $f_{sky}2Q dQ
d\phi/(2\pi)$, where the fraction of the sky in the patch is
$f_{sky}=L^2/4\pi$. The effective number of modes contributing to
$\ell$ is therefore $g_\ell = f_{sky}(2\ell+1)$, the usual $2\ell
+1 $ in the $f_{sky}$=1 all-sky case. Of course the reduction by
just $f_{sky}$ is approximate, since modes with wavenumber below
the fundamental one of $2\pi/L$ hardly contribute, inviting a more
sophisticated relation for $g_\ell$ to characterize effects from
filtering, discreteness and apodization (smooth weighting of the
target region)~\cite{boom01}. The decomposition of the
spatial-mode function in this ``momentum space" now involves a
Fourier phase factor associated with the pixel position instead of
a $Y_{\ell m}$, the beam ${\cal B}({\bf Q}) $ and the rest, $U_p
({\bf Q})$, which includes discretization into time bins and
pixelization as well as the switching-strategy and any long
wavelength filtering done in software. (In the spherical harmonic
decomposition, $U_{p,\ell m m^\prime}$ is a function of $m^\prime$
as well as the $\ell m$ and possibly the pixel position.)

Examples of $U_p ({\bf Q})$ are given in \cite{bh95}. For
switching experiments through a throw angle
${\boldsymbol\pomega}_{throw}$, $U_p ({\bf Q})$ is a first power
of $2\sin({\bf Q}{\boldsymbol \cdot} {\boldsymbol\pomega}_{throw}
/2)$ if a chop, a second power if a chop-and-wobble. A separable
form $U_p ({\bf Q})= \mu (\pomega_p \vert \pomega_C) u({\bf Q})$,
involving a spatial-mask function $\mu$ centered at some point
$\pomega_C$ and a position-independent filter $u({\bf Q})$, is
appropriate for the recent Boomerang analysis~\cite{boom01}, with
$u({\bf Q})$ a combination of temporal and spatial filters and
$\mu$ unity inside an ellipsoidal region, zero outside it.

Beams ${\cal B}_c({\bf Q}) $ for each channel $c$ must be
determined experimentally, typically through the patterns of point
source on the sky, but also through detailed ``optics" computations
of the telescope setup. Most often there is a nice monotonic
fall-off from the central point to relatively low levels of power,
though there is invariably at least some beam-asymmetry and
lurking side lobes. In the past, it was common practice to use
circularly-symmetric monotonically-falling ${\cal B}_{\ell}$'s in
CMB analysis, even ``circularizing'' pixelization effects, but now
highly accurate treatments are needed to achieve our goal of great
precision. The full width half maximum $\theta_{fwhm}$ is the
usual way to quote beam scale. If a Gaussian approximation to the
profile is appropriate, the Gaussian smoothing scale is $\ell_s
\approx 810 (10^\prime/\theta_{fwhm})$ in multipole space.

\noindent {\bf 3.4.2 Signal Correlation Matrices and Window
Functions}: For isotropic theories for which the spectra ${\cal
C}_{T\ell}$ are only functions of $\ell$, the pixel--pixel
correlation function of the signals can be expressed in terms of
$N_{pix}\times N_{pix}$ ``$\C_\ell$-window" matrices $\cW_{p
p^\prime , \ell}$:
\begin{eqnarray}
&&C_{Tpp^\prime} \equiv \avrg{\Delta_p\Delta_{p^\prime}} = {\cal
I}[ \cW_{p p^\prime , \ell}{\cal C}_{T\ell} ], \ \ \cW_{p p^\prime
, \ell}\equiv {4\pi \over 2\ell +1}  \sum_{m}
{\cal F}_{p,\ell m } {\cal F}_{p^\prime ,\ell m }^*  , \label{eq:pspec.ctpp} \\
&&{\cal I} ( f_\ell ) \equiv \sum_\ell {(\ell +{1\over 2})\over
\ell (\ell +1) } \, f_\ell\,  . \label{eq:Ilogint}
\end{eqnarray}
It is expressed in terms of ${\cal I} (f)$, the discrete
``logarithmic integral'' of a function  $f$. The average
$\overline{\cW}_{\ell}\equiv \sum_{p=1}^{N_{pix}} \cW_{p p ,
\ell}/N_{pix} $ is often used to characterize the
$\ell$-sensitivity of the experiment, since $\sqrt{{\cal I}[
\overline{\cW}_{\ell}{\cal C}_{T\ell} ]}$ gives the {\it rms}
anisotropy amplitude. For the simple case of the COBE map we have
$\cW_{p p^\prime , \ell}= \overline{\cW}_\ell P_\ell (\cos
\theta_{pp^\prime})$ in terms of the Legendre polynomials for
$\ell \ge 2$, in which both beam and pixelization effects are
taken into account in $\overline{W}_\ell $. In general, the
implicit isotropization that makes this only a function of $\ell$
does not work at high $\ell$ because of the pixelization, and this
can significantly complicate the treatment. Switching and other
spatio-temporal filters applied in software or hardware further
complicate the expressions.

\vs \noindent {\bf 3.5 Templates in Time and Space}: Templates are
specific temporal patterns that we probe for in the timestream,
$\lcU_{ct,A}$. Templates could have frequency structure ($c$) as
well. They are often associated with specific spatial patterns
$\cU_{cp,A}$ that we can probe for in the map or directly in the
timestream with $\lcU_{ct,A}$=$\sum_p P_{ctp}\cU_{cp,A}$. The
unknowns are the amplitudes $\kappa_A$ (and the errors on them),
and possibly other parameters characterizing their position,
orientation and scale. The residual signal associated with the
templates is therefore ${\cal R}_{ct} = \sum_A
\lcU_{ct,A}\kappa_A$. Templates can model contamination by
radiation from the ground, balloon or sun, components of
atmospheric fluctuations and cosmic ray hits. Often they are
synchronous with a periodic motion of the instrument.

In maps, spatial templates have been used in the removal of
offsets, gradients, dipoles, quadrupoles and looking for point
sources and extended sources (\eg IRAS/DIRBE structures). Sample
forms for $\cU_{cp}$ are: a constant for an average offset,
$\hat{q}_{pi}$ for the three dipole components,
$\hat{q}_{pi}\hat{q}_{pj}-\delta_{ij}/2$ for the five quadrupole
components, the pattern of dust emission in the IRAS/DIRBE maps,
and the beam pattern for point sources, $\cU_{cp} = {\cal B}_c
(\hat{q}_p -\hat{q}_{source})$. Even the individual pixels
themselves define spatial templates.

Since $\lcU_{ct,A}$ is of the same mathematical form as $P_{ctp}$
we can solve simultaneously for $\Delta_{cp}$ and $\kappa_A$,
delivering an extended map $\overline{\Delta}$ and $\bar{\kappa}$,
with appropriate error covariances for each, including a cross
term. The assumption is that the {\it a priori} probability for
the template amplitudes $\kappa$ is uniform, as for the map
amplitudes.

Often we just want to remove an undesired pattern from the map. We
can do this by directly projecting it from the map, and modifying
the map statistics. One can also marginalize the unobserved
$\kappa_A$, \ie integrate over them. A nice way to deal with this
is to assume that the prior probability for $\kappa$ is Gaussian
with zero mean and correlation $K_{AA^\prime}=\avrg{\kappa_A
\kappa_{A^\prime}}$. Letting the dispersion become infinite
reduces to the projection. The explicit result of marginalizing
over the template amplitudes yields modified noise weights and a
modified average map:
\begin{eqnarray}
&& \widetilde{w} \equiv w -w \lcU (\lcU^\dagger w \lcU +K^{-1}
)^{-1} \lcU^\dagger w = (w^{-1}+ \lcU K \lcU^\dagger )^{-1}  \, ,
\label{eq:modw} \\
&& \widetilde{W}_N \overline{\Delta} = P^\dagger \widetilde{w} d\,
, \qquad \widetilde{W}_N \equiv P^\dagger \widetilde{w} P \, ,
\qquad \widetilde{C}_N\equiv \widetilde{W}_N^{-1}. \nonumber
\end{eqnarray}
Letting the eigenvalues of $K$ $ \rightarrow \infty$ gives the
uniform prior, and in that case the modified $w$ has the template
vectors projected out, \ie $ \widetilde{w} \lcU$=0.

 The effect of marginalizing out $\kappa$ is therefore to enhance
the noise. One price to pay is that time-translation invariance of
the modified $w$  is lost: \ie $\widetilde{w}$ is not diagonal in
temporal frequency $\omega$, and cannot be computed directly by
the FFT, followed by smoothing. However, the relationship
eq.(\ref{eq:modw}) allows $\widetilde{c}_n\equiv
\widetilde{w}^{-1}$ to be computed with one $N_\kappa \times
N_\kappa$ matrix inversion in addition to the FFT.

\vs \noindent ${\bf 3.6 \ \cP(signals \vert parameters)}$: The
signal component of the map,
\begin{eqnarray} && \Delta_{cp} = \sum_j s_{(j)c p}  +\sum_A
\cU_{cp,A}\kappa_A \, ,
\end{eqnarray}
includes the frequency-independent primary, $s_{(1)cp}$ say,
foreground and secondary anisotropy components $s_{(j)cp}$ and
additional template-based signals $\cU_{cpA} \kappa_A$.

\noindent {\bf 3.6.1 Gaussian Signals and Wiener Filters}: For
primary anisotropies, the signal is often assumed to be Gaussian,
\begin{eqnarray}
&& \ln \cP(s \vert y) = -\half s^\dagger C_T^{-1}s - \half {\rm Tr} \ln
C_T - N_{pix}\ln \sqrt{2\pi}\, ,
\end{eqnarray}
where $C_T(y)\equiv W_T^{-1}$ is the theory pixel-pixel correlation
matrix and we have denoted the set of cosmological parameters it
depends on by $y_\alpha$. (Tr denotes trace.) A great advantage of
assuming all signals and noise to be Gaussian is that marginalizing
all signal amplitudes $\sum s_{(j)cp}$ and template amplitudes
$\kappa_A$ yields another simple Gaussian:
\begin{eqnarray}
&&\ln \cP (\overline{\Delta} \vert y) =
-\textstyle{1\over2}\overline{\Delta}^{\dagger}\widetilde{W}_t\overline{\Delta}
 +\textstyle{1\over2} {\rm Tr} \ln \widetilde{W}_t
-N_{pix}\ln \sqrt{2\pi}   \, , \label{eq:likeNC} \\
&& W_t \equiv (C_N+C_T)^{-1}\, ,  \ \widetilde{W}_t \equiv
(\widetilde{C_N}+C_T)^{-1}\, , \
\widetilde{C_N}\equiv C_N +\cU K \cU^{\dagger} \, , \nonumber \\
&& \widetilde{W}_t \equiv W_t - [\cU^{\dagger}W_t]^\dagger (K^{-1}+
\cU^{\dagger}W_t\cU)^{-1}[\cU^{\dagger}W_t] \, . \nonumber
\end{eqnarray}
As for the time templates,  the marginalization can be thought of
as inducing enhanced noise in the template spatial structures.

The other part of the probability chain, $\ln \cP (\Delta \vert
\overline{\Delta})$,  gives a Gaussian for the signal given the
observations, with mean the Wiener filter of the map:
\begin{eqnarray}
\ln \cP (\Delta \vert \overline{\Delta}) =&&
-\textstyle{1\over2}(\Delta -\avrg{\Delta \vert
\overline{\Delta}})^{\dagger}C_{\Delta \Delta}^{-1} (\Delta
-\avrg{\Delta \vert \overline{\Delta}})  \nonumber \\
&&\qquad -\textstyle{1\over2} {\rm
Tr} \ln C_{\Delta \Delta}
-N_{pix}\ln \sqrt{2\pi} ,   \nonumber \\
 \avrg{\Delta \vert \overline{\Delta}} =
C_T\widetilde{W}_t\overline{\Delta} \, ,&& \quad 
\avrg{\delta \Delta \delta \Delta^\dagger \vert
\overline{\Delta}} =C_T -C_T \widetilde{W}_tC_T \, .\label{eq:likeNC2}
\end{eqnarray}
The brackets indicate an ensemble average. The lemma
\begin{eqnarray} && B(A+B)^{-1}
= (A^{-1}+B^{-1})^{-1}A^{-1} \,
\end{eqnarray}
between two invertible matrices $A$ and $B$ is convenient to
remember. Thus the Wiener filter can also be written as
$(\widetilde{W}_N + W_T)^{-1}\widetilde{W}_N\overline{\Delta}$. 

We can also determine the distribution, given the observations, of the
various component signals $s_{(j)}$ and of the noise; all of these are
Gaussians with means and variances given by:
\begin{eqnarray} && \avrg{s_{(j)}\vert \overline{\Delta}} =
C_{T(j)} \widetilde{W}_t \overline{\Delta} , \qquad \ \avrg{\delta
s_{(j)} \delta s_{(j^\prime)}^\dagger} = C_{T(j)}
\delta_{jj^\prime}- C_{T(j)}\widetilde{W}_t
C_{T(j^\prime)} , \nonumber \\
&& \avrg{n\vert \overline{\Delta}} = C_{N} \widetilde{W}_t
\overline{\Delta} , \qquad \qquad \avrg{\delta n \delta
n^\dagger} = C_{N} - C_{N}\widetilde{W}_t
C_{N} , \nonumber \\
&&\delta s_{(j)} \equiv s_{(j)} - \avrg{s_{(j)}\vert \overline{\Delta}}
, \qquad \quad \delta n \equiv n- \avrg{n\vert
\overline{\Delta}}.\label{eq:wiener}
\end{eqnarray}

\noindent {\bf 3.6.2 Non-Gaussian Signals and Maximum Entropy
Priors}: For foregrounds and secondaries, all higher order
correlations are needed since the distributions are non-Gaussian.
One sample non-Gaussian prior distribution that has been analyzed
for these signals and even for primaries is the so-called maximum
entropy distribution, which has been widely used by radio
astronomers in image construction from interferometry data.
Although ``max-ent'' is often a catch-all phrase for finding the
maximum likelihood solution, the implementation of the method
involves a specific assumption for the nature of ${\cal P}(s_{(j)c
p}\vert {\rm theory})$.  For positive signals, $s > 0$, it is
derived as a limit of a Poisson distribution, $P_n = \mu^n
e^{-\mu}/n!$,  in the Stirling formula limit,
\begin{eqnarray}
&& \ln n! = \Gamma (1+\mu \zeta ) \approx \mu \zeta \ln (\mu \zeta)
-\mu \zeta +\ln(2\pi \mu \zeta )/2 : \nonumber \\
&& \ln \cP(s \vert y) = (-\zeta \ln (\zeta )+\zeta-1)\mu \, \ {\rm
where} \ \ \zeta =s/\mu \, .
\end{eqnarray}
The subdominant $\ln(2\pi \mu \zeta )$ term is dropped and  the
Poisson origin is largely forgotten, replaced by an interpretation
involving the classic Boltzmann entropy, $- \zeta \ln \zeta$, with
the  constraint on the average $\avrg{\zeta}=1$ and $\mu$ now
interpreted as a measure.

Another example of positive signals is standard emitting sources.
These often have power law distributions over some flux range,
with leading term $\ln \cP = -\gamma \ln \zeta$. These have to be
regulated at small and/or large $\zeta$ to converge.

For symmetric positive and negative signals, a possible form is
\begin{eqnarray}
&& {\cal S}_P (s_{(j)} ) = \ln \cP(s_{(j)} \vert y) = -x \ln
(x+\sqrt{1+x^2})+\sqrt{1+x^2}-1 , \nonumber \\
&& \ \ x^2 \equiv s_{(j)}^\dagger C_{T(j)}^{-1}s_{(j)}  ,
\end{eqnarray}
which reduces to $-x^2/2$ in the small fluctuation (small $x$)
limit, but has non-Gaussian wings. This form has been used in CMB
forecasting for Planck by \cite{lasenby00,hobson}, but any form which
retains the basic $-x^2/2$ limit for small $x$ and a levelling off
at high $x$ can have a similar effect. For example, instead of the
$-{\rm arcsinh}(x)$ for $\partial \ln P/\partial x$,  the more
drastic deviation from Gaussian, $-{\rm arctan}(x)$, has been used
in radio astronomy.

Of course, in spite of the analytic form for ${\cal S}_P$, the
integration over the various $s_{(j)}$ to get ${\cal
S}(\overline{\Delta}\vert y)$ does not have an analytic result.
Nonetheless, iterative techniques can be used to solve for the
maximum entropy solution, and errors can be estimated from the
second derivative matrix of the total entropy.

\vs \noindent {\bf 3.7 Cleaning and Separating}: Separating
foregrounds and secondaries from the primary CMB into
statistically accurate maps is a severe challenge. It has been
common to suppose $s_{(j)cp} = f_{(j)c} s_{(j)p}$ is separable
into a product of a given function of frequency times a spatial
function. But this is clearly not the case for foregrounds, and
for most secondary signals, although for some, such as the
Sunyaev-Zeldovich effect, it is nearly so
(Fig.~\ref{fig:frequencyforegrounds}). Another approach which is
crude but reasonably effective is to separate the signals using
the multifrequency data on a pixel-by-pixel basis, but the
accuracy is limited by the number of frequency channels.

\noindent {\bf 3.7.1 Gaussian {\it cf.} Maximum Entropy Component
Separation}: It is clearly better to incorporate the knowledge we
do have on the $s_{(j)cp} $ spatial patterns. This has been done
so far by either assuming the prior probability for ${\cal
P}(s_{(j)c p}\vert {\rm theory})$ is Gaussian or of the max-ent
variety, or is an amplitude times a template. Even the obviously
incorrect Gaussian approximation for the foreground prior
probabilities has been shown to be relatively effective at
recovering the signals in simulations for Planck performed by
Bouchet and Gispert \cite{bouchetgispert99}; see also
\cite{prunet00b}.

The Poisson aspect of the maximum entropy prior makes it
well-suited to find and reconstruct point sources, and more
generally concentrated ones: for Planck-motivated simulations, it
has been shown to be better at recovery than the Gaussian
approximation \cite{hobson}. Since errors are estimated from the
second derivative matrix,  non-Gaussian aspects of the errors are
ignored. Further, foregrounds and secondary anisotropies have
non-Gaussian distributions which are certainly not of the max-ent
form, and can only be determined by Monte Carlo methods,
simulating many maps
--- and then only if we know the theory well enough to construct
such simulations, a tall task for the foregrounds. Fortunately,
with current data these issues have not been critical to solve
before useful cosmological conclusions could be drawn, but much more
exploration is needed to see what should be done with separation
for the very high quality Planck and MAP datasets.

\noindent {\bf 3.7.2 Component Separation with Templates}: For
foreground removals, maps at higher or lower frequencies than the
target one, where the foregrounds clearly dominate, can be used as
the templates. The marginalized formulas given in \S 3.5, 3.6 get
rid of the templates but their amplitudes and distribution may
also be of great interest. The other part of the likelihood
(before marginalization) is $\ln \cP (\kappa \vert
\overline{\Delta})$; with the Gaussian prior assumption, it is a
Gaussian distribution with the mean a weighted template-filtering
of the data, akin to a Wiener-filtering:
\begin{eqnarray}
&& \avrg{\kappa \vert \overline{\Delta}} = [K^{-1}+
\cU{^\dagger}W_t\cU ]^{-1} \cU{^\dagger}W_t \overline{\Delta}
\rightarrow [\cU{^\dagger}W_t\cU ]^{-1} \cU{^\dagger}W_t
\overline{\Delta}\, , \nonumber \\ && C_{\kappa \kappa} = \avrg{\delta
\kappa \delta \kappa^\dagger\vert \overline{\Delta}} =
[\cU{^\dagger}W_t\cU ]^{-1}, \ \qquad \delta \kappa = \kappa -
\avrg{\kappa\vert \overline{\Delta}} .\label{eq:kappawiener}
\end{eqnarray}

An example \cite{jfb99} of the use of templates was comparing Wiener
maps and various statistical quantities for the SK95 data under the
hypothesis that it was pure CMB or CMB plus a ``dust template" -- a
cleaned IRAS $100 \mu m$ map with strong input from the DIRBE 100 and
240 $\mu m$ maps \cite{schlegel98}.  A correlated component in the
30--40 GHz SK95 data was found \cite{jfb99}, but at a low level
compared with the CMB signal, \ie not much dust contamination, in
agreement with what other methods showed \cite{deoliveracosta97}.

\noindent {\bf 3.7.3 Point Source Separation with Templates}: A
useful method for  finding localized sources is to look for known
or parameterized non-Gaussian patterns with the position, and
possibly orientation and scale of the templates being allowed to
float. An example is (optimal) point source removal, in which the
map is modelled as a profile times an amplitude $\kappa$ plus
noise plus the other signals present \cite{cb01}. As the source
position varies, eq.(\ref{eq:kappawiener}) defines a
$\overline{\kappa}$--map. The dimensionless map $\bar{\nu} \equiv
C_{\kappa\kappa}^{-1/2}\overline{\kappa}$ gives the number of
$\sigma$ a signal with the beam pattern shape has. If $\vert
\bar{\nu} \vert$ is small, \eg below $2.5\sigma,$ it will be
consistent with the Gaussian signals in the map; if large, \eg
above $5\sigma$, it will stick out as a non-Gaussian spike ripe
for removal. The question then is how many $\sigma$ one cleans to,
and what impact setting such a threshold has on the statistics of
the cleaned map, $\Delta^\prime = \Delta -\cU \avrg{\kappa \vert
\overline{\Delta}}$ with its corrected weight matrix
$\widetilde{W}_t^\prime = W_t -W_t \cU C_\kappa \cU{^\dagger}
W_t$. (Actually $\widetilde{W}_t^\prime$ always acts on $\Delta$,
so removing $\cU \avrg{\kappa \vert \overline{\Delta}}$ is not
necessary in the limit of a wide prior dispersion for $\kappa$.)

Fig.~\ref{fig:SPsig} shows an application of this method to DMR. A
comparison is also made between the mean noise field and the
$\overline{\kappa}$ field: they are quite similar.

A similar optimal filter was used in \cite{tegmark98} for ideal
forecasting and in \cite{tegmarkSZ98} for SZ source finding (based on
the beta-profile).  Another approach is to not pay such attention to
the detailed optimal filter, but use a Gaussian-based filter with
variable scale to try to identify point sources.  \cite{cayon00} used
a Mexican hat filter, \ie the second derivative of a Gaussian. This
continuum wavelet method is optimal only in the case of a scale
invariant signal, a Gaussian beam and no noise. In the pure white
noise limit with no signal, the straight Gaussian is optimal. The best
choice lies in between.

\vs \noindent {\bf 3.8 Comparing}: The simplest way to compare
data sets A and B is to interpolate on overlap regions dataset A
onto the pixels of dataset B. This of course will not always be
possible, especially if the pixels are generalized ones, \eg modal
projections such as in the SK95 dataset.  We would also like to
compare nearby regions even if overlap is only partial. This
requires an extrapolation as well as interpolation. In
\cite{knoxcompare98},  a simple mechanism was described based upon
Bayesian methods and a Gaussian interpolating theory for comparing
CMB data sets. The statistics of the two experiments is defined by
a joint probability for the two datasets, $\cP(A,B\vert C_T)$, where
the theory matrix $C_T$ has $AB$ as well as $AA$ and $BB$ parts
connecting it. A probability enhancement factor $\beta = \ln
\cP(A\vert B,C_T) /\cP(A\vert C_T)$ was introduced, which is invariant
under $A,B$ interchange.

What was visually instructive was to compare the A-data
Wiener-filtered onto B-pixels, $\avrg{s_B \vert \Delta_A} =
C_{T,BA}\widetilde{W}_{t,AA}\overline{\Delta}_A$, with the
Wiener-filter of B-data, $\avrg{s_B \vert \Delta_B} =
C_{T,BB}\widetilde{W}_{t,BB}\overline{\Delta}_B$. They should bear
a striking resemblance except for the errors. The example used in
\cite{knoxcompare98} was SK95 data for A and MSAM92 for B, which
showed remarkable similarities: indeed that the two experiments
were seeing the same sky signal, albeit with some deviations near
the endpoint of the scan. Although these extrapolations and
interpolations are somewhat sensitive to the interpolating theory,
if it is chosen to be nearly the best fit the results are robust.

The same techniques were used for an entirely different purpose in
\cite{jaffe00}, testing whether two power spectra with appropriate
errors, in this case from Boomerang and Maxima, could have been
drawn from the same underlying distribution. The conclusion was
yes.

\vs \noindent {\bf 3.9 Compressing}: We would like to represent in
a lossless way as much information as we can in much smaller
datasets.  Timestreams to maps (and map-orthogonal noise) are a
form of compression, from $N_{tbits}$ to $N_{pix}$. Map
manipulations, even  for Gaussian theories,  are made awkward
because the total weight matrix involves $C_N$, often simplest in
structure  in the pixel basis, and $C_T$, which is often naturally
represented in a spherical harmonic basis. Signal-to-noise
eigenmodes $\xi_k$ are a basis in which $\widetilde{C}_N^{-1/2}
C_T \widetilde{C}_N^{-1/2}$ is diagonal, hence are statistically
independent of each other if they are Gaussian. They are a
complete representation of the map. Another S/N basis of interest
is one in which $C_T^{1/2} \widetilde{C}_N^{-1} C_T^{1/2}$ is
diagonal, since $\widetilde{C}_N^{-1}$ comes out directly from the
signal/noise extraction step.  Finding the S/N modes  is another
$O(N_{pix}^3)$ problem. Typically, $\avrg{\xi_k^2}$ falls off
dramatically at some $k$, and higher modes of smaller eigenvalues
can be cut out without loss of information. These truncated bases
can be used to test the space of theoretical models by Bayesian
methods \eg \cite{bj99}, and determine cosmic parameters. Further
compression to bandpowers is even better for more rapid
determination of cosmic parameters.

The related compression concept of parameter eigenmodes mentioned
in \S~1.4 finds linear combinations of the cosmological or other
variables $y_\alpha$ that we are trying to determine which are
locally statistically-independent on the multidimensional
likelihood surface. It provides a nice framework for dealing with
near-degeneracies.

\vs \noindent {\bf 3.10 Power Spectra and Parameter Estimation}:
Determining the statistical distributions of any target parameters
characterizing the theories, whether they are cosmological in
origin or power amplitudes in bands or correlation function in
angular bins, is a major goal of CMB analysis. For power spectra
and correlation functions, it is natural that these would involve
pairs of pixels. Operators linear in pixel pairs are called {\em
quadratic estimators}. Maximum likelihood expressions of
parameters in Gaussian theories also often reduce to calculating
such quadratic forms, albeit as part of iteratively-convergent
sequences. Hence the study of quadratic estimators has wide
applicability.

\noindent {\bf 3.10.1 Quadratic Estimators}: Estimating power
spectra and correlation functions of the data has traditionally
been done by minimizing a $\chi^2$ expression,
\begin{eqnarray}
&&\chi^2 = Tr [W(\overline{\Delta} \overline{\Delta}^\dagger
-C(y))W (\overline{\Delta} \overline{\Delta}^\dagger -C(y))] \,
/\, 4,
\end{eqnarray}
where $W$ is some as yet unspecified weight matrix. The $1/4$ is
from the pair sum. The critical element is to make a model $C(y)$
for the pixel pair correlation $\overline{\Delta}
\overline{\Delta}^\dagger$ which is as complete a representation
as possible. For example, if we adopt a linear dependence about
some fiducial $C_*=C(y_*)$,
{i.e., } $ C(y) = C_* + \sum_\beta C_\beta \delta y^\beta $, 
then
\begin{eqnarray}
&& \sum_{\beta} {\cal F}_{\alpha \beta} \delta
y^\beta = \half {\rm Tr} [WC_\alpha W (\overline{\Delta}
\overline{\Delta}^\dagger -C_*)] , \ \ C_\alpha \equiv \partial
C/\partial y^\alpha \, , \label{eq:quadsoln}\\ &&
{\cal F}_{\alpha \beta} \equiv \half {\rm Tr} [WC_\alpha W C_\beta]
-\half {\rm Tr} [WC_{\alpha \beta} W (\overline{\Delta}
\overline{\Delta}^\dagger -C_*)] ,\ \  C_{\alpha \beta} \equiv
{\partial^2 C \over \partial y^\alpha
\partial y^\beta }, \nonumber \\ && {\rm Fisher \ information \
matrix}: \ \ F_{\alpha \beta} \equiv \avrg{{\cal F}_{\alpha
\beta}} = \half {\rm Tr}[WC_\alpha W C_\beta] \, . \label{eq:fisherdefn}
\end{eqnarray}
Since ${\cal F}_{\alpha \beta}$ has $\Delta$ dependence, this is
not strictly a quadratic estimator. However, it is if the Fisher
matrix $F_{\alpha \beta}$, the ensemble average of ${\cal
F}_{\alpha \beta}$, is used. Iterations with $F$ rather than
${\cal F}$  converge to the same $y_*$ if $y$ is continually
updated by the $\delta y$.   $C_* = C_N +C_T$ has been assumed.

The correlation in the parameter errors can be estimated from the
average of a quartic combination of pixel values,
\begin{eqnarray}
&& \avrg{\delta y \delta y^\dagger} =
{\cal F}^{-1} \half {\rm Tr}[WC_\alpha WCWC_\beta WC]{\cal F}^{-1} \nonumber \\
&& \quad \qquad + {\cal F}^{-1} {\textstyle{1\over 4}}
 {\rm Tr}[WC_\alpha W(\delta C) ]{\rm Tr}[WC_\beta W(\delta C) ]{\cal F}^{-1} \, ,  \nonumber 
\end{eqnarray}
where  $(\delta C) \equiv C -C_* .$
Note,  $\avrg{\delta y \delta y^\dagger} = {\cal F}^{-1}$ if 
$C= C_*$ and $ \ W=C_*^{-1} . $ 

Another approach is to estimate the errors by taking the second
derivative of $\chi^2$, $ \partial
\chi^2 /\partial y^\alpha \partial y^\beta$  = ${\cal F}_{\alpha \beta}$:
\begin{eqnarray}
 \avrg{\partial^2 \chi^2 /\partial y _\alpha \partial
y_\beta} &&=\half {\rm Tr}[WC_\alpha WC_\beta ]
- \half {\rm Tr} [WC_{\alpha \beta}W (\delta
C)]  \nonumber \\
&& \rightarrow F_{\alpha \beta} \ {\rm as} \ (\delta C)
\rightarrow 0 . \nonumber
\end{eqnarray}
These two limiting cases hold for the maximum likelihood solution
as we now show.

\noindent {\bf 3.10.2 Maximum Likelihood Estimators \& Iterative
Quadratics}: The maximum likelihood solution for a Gaussian signal
plus noise is of the form eq.(\ref{eq:quadsoln}) with
$W=\widetilde{W}_t =(\widetilde{C}_N+C_T(y))^{-1}$, the ``optimal"
weight which gives the minimum error bars. This is found by
differentiating $-2\ln {\cal L} = {\rm Tr}[\widetilde{W}_t
\overline{\Delta} \overline{\Delta}^\dagger)]-{\rm
Tr}[\ln\widetilde{W}_t]+N_{pix}\ln(2\pi)$. Of course the $y_\beta$
dependence of $\widetilde{W}_t$ means the expression is not really
a quadratic estimator. A solution procedure is the Newton-Raphson
method, with the weight matrix updated in each iterative
improvement $\delta y_\beta$ to $y_{*\beta}$, at a cost of a
matrix inversion. At each step we are doing a quadratic
estimation, and when the final state has been converged upon, the
weight matrix can be considered as fixed.  Solving for the roots
of $\partial \ln{\cal L}/\partial y_\alpha$ using the
Newton-Raphson method requires that we calculate $\partial^2
\ln{\cal L}/\partial y_\alpha \partial y_\beta$, the curvature of
the likelihood function. Other matrices have been used for
expediency, since one still converges on the maximum likelihood
solution; in particular  the curvature matrix expectation value,
\ie the Fisher matrix $F_{\alpha \beta}$, which is easier to
calculate than ${\cal F}_{\alpha \beta}$. Further, to ensure
stability of the iterative Newton-Raphson method, some care must
be taken to control the step size each time the $y_\beta$ are
updated. Since the $\avrg{\delta y_\alpha \delta y_\beta}$
estimate is just $[{\cal F}^{-1}]_{\alpha \beta}$, this matrix must at
least be determined after convergence to characterize the
correlated errors.

 Note that only in the Gaussian and uniform prior cases is the integration
over $s_{(j)}$ analytically calculable, although maximum
likelihood equations can be written for simple forms that can be
solved iteratively, \eg maximum entropy priors.

\noindent {\bf 3.10.3 The Bandpower Case}: The  bandpower
associated with a given window function $\varphi_{b\ell}$ of a
theory with spectrum ${\cal C}_{T\ell}$ is defined as the average
power, 
\begin{equation}
\avrg{{\cal C}_\ell}_{b}\equiv {\cal I}[ \varphi_{b\ell}{\cal
C}_{T\ell} ]/{\cal I}[ \varphi_{b\ell}] , \label{eq:bandpow}
\end{equation}
with ${\cal I}$ defined by eq.(\ref{eq:Ilogint}). Many choices are
possible for $\varphi_{b\ell}$, but the simplest is probably the
best: \eg $\chi_b (\ell )$ which is unity in the $\ell$-band, zero
outside. Another example uses the average window function of the
experiment, $\overline{\cW}_{\ell}\chi_b (\ell )$; others use
window functions related to relative amounts of signal and noise.
There is some ambiguity in the choice for the filter
$\varphi_{b\ell}$ ~\cite{bjk00,knox00}, but often not very much
sensitivity to its specific form.

We want to estimate bandpowers $q^b$ taken relative to a general
shape ${\cal C}^{(s)}_{\ell}$ rather than the flat shape: $q^b
\equiv {\cal I} (\varphi_{b\ell} {\cal C}_{T\ell} )/ {\cal I}
(\varphi_{b\ell} {\cal C}^{(s)}_\ell )$, so $\avrg{{\cal C}}_b =
\bar{q}_b \avrg{{\cal C}^{(s)}}_b$. We therefore model
$\overline{\Delta} \overline{\Delta}^\dagger$ by
\begin{eqnarray}
&&
C_{pp^\prime}(q)= \widetilde{C}_{N,pp^\prime}+\sum_b
C^{(s)}_{b,pp^\prime} q^b \equiv C_{*,pp^\prime} + \sum_b
C^{(s)}_{b,pp^\prime}\delta q^b , \nonumber \\
&& C^{(s)}_{b,pp^\prime} =
{\cal I}[ \cW_{p p^\prime , \ell}\chi_b (\ell ) {\cal C}_{\ell}^{(s)}]\,
, \qquad \delta q^b=q^b-q^b_*\, .
\end{eqnarray}
Here the $q^b_*$ are the bandpowers on a last iteration,
ultimately converging to maximum likelihood bandpowers if the
relaxation is allowed to go to completion, $\delta q^b \rightarrow
0$. The pixel-pixel correlation matrices
$C_{Tb,pp^\prime}=C^{(s)}_{b,pp^\prime}q^b $ for the bandpowers
$b$ follow from eq.(\ref{eq:pspec.ctpp}).

The usual technique \cite{bjk98} is to use this linear expansion in
the $q^b$, the downside being that the bandpowers could be negative --
when there is little signal and the pixel pair estimated signal plus
noise is actually less than the noise as estimated from $C_N$.  By
choosing $\exp (q^b)$ rather than $q^b$ as the variables, positivity
of the signal bandpowers can be assured. (The summation convention for
repeated indices is often used here.)

The full maximum likelihood calculation has been  used for
bandpower estimates for COBE, SK95, Boomerang and Maxima, among
others. The cost is ${\cal O}(N_{pix}^3)$ for each iteration, so
this limits the number of pixels that can be treated. For $N_{pix}
\gta 2\times 10^5$ or so, the required computing power becomes
prohibitive, requiring 640~Gb of memory and of order
$3\times10^{17}$ floating-point operations, which translates to a
week or more even spread over a Cray T3E with $\sim1024$ 900-MHz
processors. This will clearly be impossible for the ten million
pixel Planck data set. Numerical roundoff would be prohibitive
anyway.

The compressed bandpowers estimated from all current data and the
satellite forecasts shown in Fig.~\ref{fig:newtnatocl9}
were also computed using the maximum likelihood Newton-Raphson
iteration method (for rather fewer ``pixels").

\noindent {\bf 3.10.4 A Wavenumber-basis Example}: There is an
instructive case in which to unravel the terms of
eq.(\ref{eq:quadsoln}). If we assume the noise and signal-shape
correlation matrices in the wavenumber basis ${\bf Q}$ of \S~3.4,
$\widetilde{C}_{N,{\bf Q}{\bf Q^\prime}}$ and $C^{(s)}_{Tb,{\bf
Q}{\bf Q^\prime}}$,  are functions only of the magnitude of the
wavenumber, $Q\approx \ell +\half$, then
\begin{eqnarray}
&& \sum_{b^\prime} F_{bb^\prime}q^{b^\prime}=  \half \sum_{Q} g_Q W^2 (Q)
C_{Tb}^{(s)} (Q) (C_{est}(Q)-\widetilde{C}_{N}(Q))
\, ,\label{eq:qbisonoise} \\
&& F_{bb^\prime} =  \half \sum_{Q}g_Q W^2 (Q)
C^{(s)}_{Tb}( Q )C^{(s)}_{Tb^\prime}( Q ) \, ,\nonumber \\
&& W(Q)=(\widetilde{C}_N(Q)+\sum_b C^{(s)}_{Tb}(Q) q_*^b)^{-1}\, \quad
C_{est}(Q)\equiv \int {d\phi_{\bf
Q} \over 2\pi}\, \vert \overline{\Delta}_{\bf Q} \vert^2 \, . \nonumber
\end{eqnarray}
Here $g_Q$ is the effective number of modes contributing to the
$\Delta Q$=1 wavenumber interval.

In the all-sky homogeneous noise limit, this of course gives the
correct result, with $g_Q=f_{sky}(2\ell+1)$ and
$C^{(s)}_{Tb}(Q)=C^{(s)}_\ell \overline{\cW}_\ell \chi_b (\ell)$,
where $\overline{\cW}_\ell$ is the  window (the angle-average of
$\vert u({\bf Q})\vert^2 {\cal B}_\ell^2$ in the language of
\S~3.4.1). Since $\chi_b (\ell) \chi_{b^\prime} (\ell)=
\delta_{bb^\prime} (\ell)\chi_b (\ell) $, $F_{bb^\prime}$ is diagonal, with a
value expressed in terms of a band-window $\varphi_{b\ell}$:
\begin{eqnarray}
&&F_{bb}=\half {\cal I}[\varphi_{b\ell}{\cal C}^{(s)}_\ell ]  ,
\quad \varphi_{b\ell}= 4\pi (g_Q/(2Q))
W^2(Q)C_\ell^{(s)}\overline{\cW}_\ell^2 \chi_b(\ell) \,
.\label{eq:SNwindow}
\end{eqnarray}
It may now seem that we have identified up to a normalization the
correct band-window to use in this idealized case, with a
combination of  signal and noise weighting. However the ambiguity
referred to earlier remains, for what we are actually doing in
estimating the observed $q^b$ is building a piecewise
discontinuous model relative to the shape ${\cal C}^{(s)}_\ell$.
The analogue  for a given ${\cal C}_{T\ell}$ is the model
$\sum_\ell \chi_b(\ell) q_{Tb}{\cal C}^{(s)}_\ell$. Since for the
theory, $q^b = {\cal I}[\varphi_{b\ell} {\cal C}_{T\ell}]/ {\cal
I}[\varphi_{b\ell} {\cal C}^{(s)}_{\ell}]$, any $\varphi_{b\ell} $
satisfying $\varphi_{b\ell}\chi_{b^\prime}(\ell)$ =
$\varphi_{b\ell}$ will do to have $q_{Tb}$ the ratio of
target-bandpower to shape-bandpower.

The downweighting by signal-plus-noise power is just what is
required to give the correct variance in the signal power estimate
in the limit of small bin size:
\begin{eqnarray}
&&\avrg{(\Delta {\cal
C}_{Tb})^2} = F_{bb}^{-1}\avrg{{\cal C}_\ell^{(s)} }_b^2   \rightarrow
{2\over g_\ell}{(
{\cal C}_{T\ell}\overline{\cW}_\ell + \widetilde{{\cal C}}_{N\ell})^2 \over
\overline{\cW}_\ell^{2}} \ {\rm as } \ \Delta \ell \rightarrow 1
\,
.\label{eq:variance}
\end{eqnarray}
In the sample-variance dominated regime,
the error is ${\cal C}_{T\ell}\sqrt{2/g_\ell}$, while in the noise-dominated regime
it is $\widetilde{{\cal C}}_{N\ell}\sqrt{2/g_\ell}/
\overline{\cW}_\ell$, picking up
enormously above the Gaussian beam scale.
Recent experiments are designed to have signal-to-noise near unity on the beam scale,
hence are sample-variance dominated for lower $\ell$. The interpretation of the
$(\widetilde{C}_{N\ell}+ C_{T\ell})^{-1}$
weighting factor is that correlated
spatial patterns associated with a wavenumber $\ell$ that are statistically
high just because of sample variance are to be downweighted by $C_{T\ell}$.

Although these results are only rigorous for the all-sky case with
homogeneous and isotropic noise, they have been used to great effect
for forecasting errors  (\S 3.11) for $f_{sky} < 1$ regions.

\noindent {\bf 3.10.5 Faster Bandpowers}: Since the need
for speed to make the larger datasets tractable is essential, much
attention is now being paid to much faster methods for
bandpower estimation.

A case has been made for highly simplified
weights (\eg diagonal in pixel space) that would still deliver the
basic results, with only slightly increased error bars
\cite{szapudi00}. The use of non-optimal weighting schemes has had a long history.
For example, a quadratic estimator of the correlation function was
used in the very first COBE DMR detection publication, with
$diag(W_N)$ for $W$, a choice we have also used for COBE and the
balloon-borne FIRS. Here one is estimating correlation function
amplitudes $C^\alpha$, where $\alpha$ refers to an angular bin:
$C_{pp^\prime}=C_{Npp^\prime}+\sum_\alpha C^\alpha \chi_\alpha
(\theta_{pp^\prime}) $, where $\cos (\theta_{pp^\prime}) =\hq_p
\cdot \hq_{p^\prime}$ and $\chi_\alpha (\theta )$ is one inside
and zero outside of the bin. As for bandpowers, the distribution
is non-Gaussian, so errors were estimated using Monte Carlo
simulations of the maps.

Any weight matrix $W$ which is diagonal in pixel space is easy and
fast to calculate.  In \cite{szapudi00} it was shown that making
$W$ the identity was adequate for Boomerang-like data. The actual
approach used a fast highly binned estimate of $C(\theta)$, which
can be computed quickly, which was then lightly smoothed, and
finally a Gauss-Legendre integration of $C(\theta)$ was done to
estimate $\avrg{{\cal C}}_b$.

One can still work with the full
$W=\widetilde{W}_t$ if the noise weight matrix $C_N^{-1}$ has a
simple form in the pixel basis. For the MAP satellite, it is nearly diagonal,
and a method exploiting this has been applied to MAP
simulations to good effect~\cite{osh}.

Since many of the
signals are most simply described in multipole space, it is
natural to try to exploit this basis, especially in the
sample-variance dominated limit.
In the Boomerang analysis of \cite{boom01,master01}, filtering of form
$\mu (\pomega_p ) u({\bf Q}){\cal B}(Q)$ was imposed. Many of the features
of the diagonal ansatz of \S~3.10.4 in wavenumber space were exploited,
complicated of course by the spatial mask $\mu$, which leads to mode-mode coupling.
The main approximation was to use the masked transforms for
the basic isotropic $Q$ functions, expressing where needed the
masked power spectra as linear transformations on the true underlying power spectrum
in $\ell$-space that we are trying to find. Monte Carlo
simulations were used to evaluate a number of the
terms. Details are given in \cite{boom01,master01} and elsewhere.

More could be said about making power spectrum estimation nearly
optimal and also tractable, and much remains to be explored.

\noindent {\bf 3.10.6 Relating Bandpowers to Cosmic Parameters}:
To make use of the $q^b$ for cosmological parameter estimation, we
need to know the entire likelihood function for $q^b$, and not
just assume a Gaussian form using the curvature matrix. This can
only be done by full calculation, though there are two analytic
approximations that have have been shown to fit the one-point
distributions quite well in all the cases tried~\cite{bjk00}. The
simplest and most often used is an ``offset lognormal''
distribution, a Gaussian in $z^b=\ln (q^b +q_N^b)$, where the
offset $q_N^b$ is related to the noise in the experiment:
\begin{eqnarray}
&&{\cal P}(q) \propto \exp[-\half (z-\bar{z})^b
{\cal F}^{(z)}_{bb^\prime} (z-\bar{z})^{b^\prime}], \nonumber \\
&& {\cal
F}^{(z)}_{bb^\prime} =(\bar{q}^b+q_N^b) {\cal
F}^{(q)}_{bb^\prime}(\bar{q}^{b^\prime}+q_N^{b^\prime}), \nonumber \\
&& \bar{q}^b/q_N^b = [{\cal
F}^{(q)0}_{bb^\prime}/{\cal F}^{(q)}_{bb^\prime} ]^{1/2}-1 \, ,  \label{eq:xb}
\end{eqnarray}
where $\bar{q}^b$ is the maximum likelihood value. To evaluate $q_N^b$
in eq.(\ref{eq:xb}), the curvature matrix evaluated in the absence of
signal, ${\cal F}^{(q)0}_{bb^\prime}$, is needed. For Boomerang and
Maxima analysis, the two Fisher matrices are standard output of the
``MADCAP'' maximum likelihood power spectrum estimation code
\cite{borill00}. Another (better) approach if the likelihood functions
${\cal L}(\{ q^b \})$ are available for each $q^b$ is to fit $\ln
{\cal L}$ by $-\half [(z-\bar{z})_b]^2g_b^2$, and use
$g_bG_{bb^\prime} g_{b^\prime}$ in place of ${\cal
F}_{bb^\prime}^{(z)}$, where $G_{bb^\prime}= diag({\cal
F}^{-1})^{1/2} {\cal F} diag({\cal F}^{-1})^{1/2} $. Here $ diag()$
denotes the diagonal part of the matrix in question.
Values of $q_N^b$ and other data for a variety of
experiments are given in \cite{bjk00} (where it is called $x^b$).
It is also possible to estimate $q_N^b$ using a quadratic estimator~\cite{boom01}.
To compare a given theory with spectrum ${\cal C}_{T\ell}$ with
the data using eq.(\ref{eq:xb}), the $q^b$'s need to be evaluated
with a specific choice for $\varphi_{b\ell}$. Although this should be band-limited,
$\propto \chi_{b\ell}$, as described in \S~3.10.4,
the case can be made for a number of alternate forms.

\vs \noindent {\bf 3.11 Forecasting}: A typical forecast involves
making large simplifications to the full statistical problem,
such as those used in \S~3.10.4. For
example, the noise is homogeneous, possibly white, and the fluctuations
are Gaussian. This was applied to estimates of how well cosmic
parameters could be determined for the LDBs and MAP and Planck in
the 9 parameter case described earlier \cite{bet}, and for cases
where the power spectrum parameters are allowed to open up beyond
just tilts and amplitudes to parameterized shapes
\cite{taruninflatn}. The methods were also applied when CMB
polarization was included, and when LSS and supernovae were
included \cite{eisenstein98}.

The procedure is exactly like that for maximum likelihood parameter
estimation, except one often does not bother with a realization of the
power spectrum, rather uses the average (input) value and estimates
errors from the ensemble average curvature, \ie the Fisher matrix:
$F^{-1}$ is therefore a measure of $\avrg{\delta y_\alpha \delta
y_\beta}$, where $\delta y_\alpha = y_\alpha-\bar{y}_\alpha$ is the
fluctuation of $y_\alpha$ about its most probable value,
$\bar{y}_\alpha$. As mentioned above, when the $y_\alpha$ are the
bandpowers instead of the cosmic parameters, we get the forecasts of
power spectra and their errors, as for MAP and Planck in
Fig.~\ref{fig:newtnatocl9}. Those results ignored foregrounds. However
other authors have treated foregrounds with the Gaussian and max-ent
prior assumptions in this simplified noise case
\cite{bouchetgispert99,hobson}.

Clearly forecasting can become more and more sophisticated as
realistic noise and other signals are added, ultimately to grow
into a full pipeline-testing simulation of the experiment, which all
of the CMB teams strive to do to validate their operations.

\vs \noindent {\bf Challenge}: The lesson of the CMB
analysis done to date is that almost all of the time and debate is
spent on the path to a believable primary power spectrum, with the
cosmological parameters quickly dropping out once the spectrum has
been agreed upon. The analysis challenge, requiring clever new
algorithms, gets more difficult as we move to more LDBs and
interferometers, to MAP and to Planck. This is especially so with
the ramping-up efforts on primary polarization signals, expected
at $<$10\% of the total anisotropies, and the growing
necessity to fully confront secondary anisotropies and foregrounds
simultaneously with the primary signals.

\vs \noindent {\bf Acknowledgments}: 
Many people have worked with us at CITA on aspects of the CMB analysis
touched on here, Carlo Contaldi, Andrew Jaffe, Lloyd Knox, Steve
Myers, Barth Netterfield, Ue-Li Pen, Dmitry Pogosyan, Simon Prunet,
Marcelo Ruetalo, Kris Sigurdson, Tarun Souradeep, Istvan Szapudi, and
James Wadsley, along with Julian Borrill, Eric Hivon and other members
of the Boomerang and CBI teams, and George Efstathiou and Neil Turok
at Cambridge.

\def\prd{{Phys.~Rev.~D}}
\def\prl{{Phys.~Rev.~Lett.}}
\def\apj{{Ap.~J.}}
\def\apjl{{Ap.~J.~Lett.}}
\def\apjsuppl{{Ap.~J.~Supp.}}
\def\mnras{{M.N.R.A.S.}}

\end{document}